\documentclass[3p, twocolumn, sort&compress]{elsarticle} %

\biboptions{}

\usepackage{dcolumn}%
\usepackage{bm}%
\usepackage{color}
\usepackage{amsmath}
\usepackage{amssymb}
\usepackage{booktabs}
\usepackage{siunitx}
\usepackage[flushleft, para]{threeparttable}
\usepackage{etoolbox}

\usepackage{color}
\usepackage[pdftex,colorlinks=true]{hyperref}

\newcommand{\kevnr}{\ensuremath{\textrm{keV}_\textrm{nr}}} %
\newcommand{\evnr}{\ensuremath{\textrm{eV}_\textrm{nr}}} %
\newcommand{\kevee}{\ensuremath{\textrm{keV}_{\textrm{ee}}}} %

\newcommand{\bigo}{\ensuremath{\mathcal{O}}}

\newcommand{\leff}{\ensuremath{\mathcal{L}_{\textrm{eff}}}} %
\newcommand{\qy}{\ensuremath{Q_{y}}} %
\newcommand{\ly}{\ensuremath{L_{y}}} %

\newcommand{\dd}{$\mbox{D-D}$} %
\newcommand{\dt}{$\mbox{D-T}$} %

\newcommand{\insitu}{\textit{in situ}}
\newcommand{\exsitu}{\textit{ex situ}}

\journal{Elsevier}

\makeatletter
\def\ps@pprintTitle{%
 \let\@oddhead\@empty
 \let\@evenhead\@empty
 \def\@oddfoot{}%
 \let\@evenfoot\@oddfoot}
\makeatother

\begin{document}

\begin{frontmatter}

    \title{Proposed low-energy absolute calibration of nuclear recoils in a dual-phase noble element TPC using \dd{} neutron scattering kinematics}

    \author[]{J.R.~Verbus\corref{cor1}}
    \ead{james\_verbus@alumni.brown.edu}
    \cortext[cor1]{Corresponding author}

    \author[]{C.A.~Rhyne}
    \author[]{D.C.~Malling}
    \author[]{M.~Genecov}
    \author[]{S.~Ghosh}
    \author[]{A.G.~Moskowitz}
    \author[]{S.~Chan}
    \author[]{J.J.~Chapman}
    \author[]{L.~de~Viveiros}
    \author[]{C.H.~Faham}
    \author[]{S.~Fiorucci}
    \author[]{D.Q.~Huang}
    \author[]{M.~Pangilinan}
    \author[]{W.C.~Taylor}
    \author[]{R.J.~Gaitskell}

    \address{Brown University, 182 Hope St., Providence, RI, USA}

    \begin{abstract}

    \noindent
        We propose a new technique for the calibration of nuclear recoils in large noble element dual-phase time projection chambers used to search for WIMP dark matter in the local galactic halo. 
        This technique provides an \insitu{} measurement of the low-energy nuclear recoil response of the target media using the measured scattering angle between multiple neutron interactions within the detector volume.
        The low-energy reach and reduced systematics of this calibration have particular significance for the low-mass WIMP sensitivity of several leading dark matter experiments.
        Multiple strategies for improving this calibration technique are discussed, including the creation of a new type of quasi-monoenergetic 272~keV neutron source.
        We report results from a time-of-flight-based measurement of the neutron energy spectrum produced by an Adelphi Technology, Inc.\ DD108 neutron generator, confirming its suitability for the proposed nuclear recoil calibration.
    \end{abstract}


\end{frontmatter}

\section{Introduction} \label{sec:introduction}

Dark matter experiments using liquid noble detector media have placed the most stringent limits on the spin-independent WIMP-nucleon cross-section over the majority of the WIMP mass range spanning 1--1000~GeV/c$^{2}$~\cite{AkeribAraujoBaiEtAl2015}.
Calibration of the nuclear recoil signal response of the target media over the recoil energy range expected for WIMP interactions is required to understand detector efficiency for the observation of potential dark matter events. 
The sensitivity of these experiments to low-mass WIMPs of mass $<$10~GeV/c$^{2}$ is strongly dependent upon the nuclear recoil response for low-energy nuclear recoils.
The low-mass WIMP signal interpretations of several recent dark matter experiments~\cite{Aalseth2013, Angloher2012, Bernabei2008} are in tension with recent exclusion limits placed by liquid xenon dark matter experiments~\cite{AkeribAraujoBaiEtAl2015, Aprile2012a}.
This tension reinforces the need for new low-energy, high-precision calibration of the nuclear recoil signal response in liquid noble detectors. 

Dual-phase liquid noble time projection chambers (TPCs) detect both the scintillation and ionization resulting from a particle interaction in the target media. 
The most common type of TPC used in the dark matter field uses photomultiplier tubes (PMTs) to record both the scintillation and ionization signals. 
The scintillation signal (S1) is promptly detected by PMTs lining the top and bottom of the detector's active region.
The ionization signal is produced by electrons that drift to the liquid noble target surface under the influence of an applied electric field $E_{d}$.
The electrons are extracted into the gas phase via an electric field $E_{e}$, where they produce secondary scintillation light (S2) via electroluminescence. 

We define the single quanta gain values relating the number of scintillation photons and ionization electrons to the corresponding observed number of detected photons as $g_1$ and $g_2$.
The variables $g_1$ and $g_2$ have units of detected-photons-per-scintillation-photon and detected-photons-per-ionization-electron, respectively. 
Due to the difficulty associated with precisely calibrating the detector-specific $g_1$ value, the scintillation yield is traditionally reported in terms of \leff{}, the measured scintillation yield relative to a monoenergetic electron recoil standard candle often provided by $^{57}$Co or $^{83\textrm{m}}$Kr. 
Recent large liquid noble detectors have precisely measured both $g_1$ and $g_2$ simultaneously using the anti-correlation of S1 and S2 signals~\cite{AkeribAraujoBaiEtAl2015, Akeribothers2016}.
This allows the \insitu{} calibration of both the light (\ly{}) and charge (\qy{}) yields for nuclear recoils in the absolute units of photons/\kevnr{} and electrons/\kevnr{}, respectively. 
In this paper, we use the units \kevnr{} (\kevee{}) to indicate energy deposited in the form of nuclear (electronic) recoils.

Dark matter experiments have traditionally used a continuum neutron source placed adjacent to the detector's active region to obtain an \insitu{} nuclear recoil calibration.
Frequently used calibration sources include $^{252}$Cf and $^{241}$Am/Be, which are spontaneous fission and ($\alpha$,\,n) sources, respectively.
These sources emit a continuous spectrum of neutrons with energies extending up to $\sim$10~MeV, and produce a relatively featureless recoil spectrum in the energy region of interest for WIMP searches.
The large, high-energy gamma ray to neutron ratio of these sources creates unwanted electromagnetic contamination during TPC calibrations. 
The emitted gamma ray to neutron ratio is ${\sim} 2$ and 0.6 for $^{252}$Cf and $^{241}$Am/Be, respectively~\cite{Knoll2000, Liu2007}.
The energy of these gamma rays is typically in the range 1--10~MeV~\cite{Hotzel1996, Murata2014}.
In the case of $^{241}$Am/Be, the ratio here is calculated for the 4.4~MeV gamma rays that are produced by the excited $^{12}$C state remaining after the $^{9}$Be($\alpha$,\,n)$^{12}$C reaction.\footnote{The rate of 60~keV gamma rays is much higher relative to the $^{241}$Am/Be neutron output: for 10$^{6}$ primary alpha particles from the $^{241}$Am decays, only 70~neutrons are emitted~\cite{Knoll2000}. The dominant gamma ray emission from $^{241}$Am alpha decays is this coincident 60~keV gamma ray; these can be more easily screened out in practice due to their low energy.}
Recently, photoneutron sources such as $^{88}$Y/Be have been used for low-energy nuclear recoil calibrations in various dark matter search technologies using the feature presented by the recoil spectrum endpoint~\cite{Collar2013}.
The ratio of gamma rays to neutrons produced by a typical $^{88}$Y/Be source is $\sim$4$\times10^{5}$ to 1~\cite{Knoll2000}.
High-Z shielding several 10~cm thick surrounding such a source is required to reduce the gamma ray rate to manageable levels for a nuclear recoil calibration.
Extracting the signal yields using such a source requires a Monte Carlo simulation, which includes a model of the initial neutron energy spectrum produced by the source and calculation of neutron energy loss in passive shielding and detector materials. 
Extraction of the nuclear recoil signal yields using these sources requires precise modeling of the source neutron spectrum and scattering inside passive detector materials to create a best-fit Monte Carlo simulation comparison to the observed recoil spectrum~\cite{Sorensen2009, Horn2011, AprileAlfonsiArisakaEtAl2013}.
The energy scale in these methods is often left as a free parameter in the overall fit to the observed signal spectra.

Existing calibrations using a fixed scattering angle to set an absolute energy scale have focused on \exsitu{} calibrations using liquid noble test cells~\cite{AprileBaudisChoiEtAl2009, Manzur2010, Plante2011, CaoAlexanderAprahamianEtAl2015}.
In these experiments monoenergetic neutrons with a known direction interact in a small liquid noble detector.
Coincident pulses in a far secondary detector are used to tag valid events.
The neutron source and detector geometry is arranged to enforce a known fixed scattering angle in the liquid noble target media.
The recoil energy $E_{\textrm{nr},A}$ is determined by Eq.~\ref{eq:recoil_energy_equation}, where $m_{A}$ is atomic mass of the target element, $E_{n}$ is the incident energy of the neutron, $m_{n}$ is the mass of the neutron, and $\theta_{\textrm{CM}}$ is the scattering angle in the center-of-mass frame:

\begin{equation} \label{eq:recoil_energy_equation}
    E_{\textrm{nr},A} = \zeta E_{n} \, \text{,}
\end{equation}

\noindent
where

\begin{equation} \label{eq:tof_zeta}
    \zeta = \frac{4 m_{n} m_{A}}{\left(m_{n} + m_{A}\right)^{2}} \frac{\left(1 - \cos{\theta_{\textrm{CM}}}\right)}{2}  \, \text{.}
\end{equation}

\noindent
The relationship between $\theta_{\textrm{CM}}$ and the scattering angle in the laboratory frame, $\theta_{\textrm{lab}}$, is given by:

\begin{equation} \label{eq:recoil_angle_cm_to_lab}
    \tan{\theta_{\textrm{lab}}} = \frac{\sin{\theta_{\textrm{CM}}}}{m_{n}/m_{A} + \cos{\theta_{\textrm{CM}}}} \, \text{.}
\end{equation}

\noindent
For target elements with large atomic mass, the approximation $\theta_{\textrm{CM}} \approx \theta_{\textrm{lab}}$ is often made, and Eq.~\ref{eq:recoil_energy_equation} can be used directly.
The maximum error in recoil energy when using approximation for argon and xenon target nuclei is 5\% and 1.5\%, respectively.
This error is determined by comparing the recoil energy in Eq.~\ref{eq:recoil_energy_equation} when using the exact value of $\theta_{\textrm{CM}}$ to that calculated using the approximation $\theta_{\textrm{CM}} \approx \theta_{\textrm{lab}}$.

These \exsitu{} calibrations can suffer from several undesirable background contributions.
First, neutrons can scatter in passive materials either before or after interacting in the liquid noble test cell, and then subsequently complete the journey to the far detector.
These neutrons lose an undetermined amount of energy during their scatters in passive material and have a poorly defined scattering angle in the liquid noble test chamber.
These effects make inference of the deposited nuclear recoil energy in the target medium difficult.
Neutrons that scatter in passive materials during their journey between the liquid xenon cell and the far detector provide a similar source of background events.
Second, it is difficult to differentiate events consisting of multiple elastic scatters in the liquid noble target during single-phase operation as is typically used for \exsitu{} \ly{} studies.
These multiple elastic scatter events will have a systematic increase in the observed scintillation signal and a measured scattering angle that is no longer directly related to the path taken through the liquid noble target.
Finally, due to the physical size of the detectors, there is a systematic uncertainty associated with the range of allowed scattering angles.
It is possible to attempt to accommodate these effects on average and estimate the associated systematic uncertainties using a neutron transport Monte Carlo simulation with a model of the experimental setup, but a more direct calibration technique can eliminate these systematic uncertainties entirely.

We present a new scattering-angle-based technique for an \insitu{}, absolute nuclear recoil calibration in modern, large, liquid-noble-based TPCs used for rare event searches~\cite{Aprile2012, AlexanderAltonArisakaEtAl2013, Akerib2014, AkeribAkerlofAkimovEtAl2015}.
In this technique, neutrons of known energy and direction are fired into a large liquid noble TPC~\cite{Gaitskell2008}.
The detector's position reconstruction capabilities provide the ($x$,\,$y$,\,$z$) coordinates of each interaction in multiple-scatter events.
The calculated scattering angle provides a direct measurement of the recoil energy at each scattering vertex according to Eq.~\ref{eq:recoil_energy_equation}.

An ideal neutron source for this type of measurement should have the following characteristics:
\begin{itemize}
    \item The neutron source should be compact and portable to allow deployment in deep underground laboratory space.
    \item In order to precisely define $E_{n}$, the neutron source source must produce a monoenergetic energy spectrum, ideally with a width ($\sigma/\mu$) subdominant to other systematic effects contributing to spectrum broadening described in Sec.~\ref{sec:dd_proposal}.
    \item To calibrate noble gas detectors in the nuclear recoil energy region of interest, the techniques described in this paper require an incident neutron beam with a mean energy between 100~keV and several MeV.
    \item The total flux into $4\pi$ solid angle of the neutron source should be greater than $\sim$10$^{7}$~n/s to achieve useful calibration rates using the technique described in Sec.~\ref{sec:dd_proposal}.
        A flux of $\sim$10$^{9}$~n/s or greater is advantageous for the creation of a 272~keV neutron reflector source as described in Sec.~\ref{sec:d_backscatter}. 
    \item The ability to pulse the neutron beam provides several advantages.
        First, controlling the duty cycle provides a precise tuning mechanism for the neutron yield.
        Second, the known ``beam on'' time during low duty cycle operation can provide a powerful reduction in calibration backgrounds.
        Third, if neutron bunch widths of $\lesssim$10~$\mu$s are achievable, then more sensitive measurement techniques described in Sec.~\ref{sec:advanced_dd_techniques} become feasible.
\end{itemize}

Several candidate monoenergetic neutron sources are available that provide required energy, flux, and pulsing characteristics.
The endothermic $^{7}$Li(p,\,n)$^{7}$Be reaction has a Q value of $-1.644$~MeV~\cite{Csikai1987}.
This reaction can provide a source of monoenergetic neutrons of tunable mean energy by accelerating the incident protons to a fixed energy above the reaction threshold.
A dedicated proton accelerator facility is required to generate the $\sim$2~MeV protons used for this reaction. 
A number of recent \exsitu{} nuclear recoil calibrations have made use of such facilities~\cite{Joshi2014, JoshiSangiorgioBernsteinEtAl2014, CaoAlexanderAprahamianEtAl2015}. 
The exothermic $^{2}$H(d,\,n)$^{3}$He (\dd{}) and $^{3}$H(d,\,n)$^{4}$He (\dt{}) reactions have Q values of 3.269~MeV and 17.590~MeV, respectively~\cite{Csikai1987}.
The modest 100~kV potential typically used to accelerate deuterium ions used for these reactions can be easily generated via compact, commercially available high-voltage supplies. 
It is typically possible to achieve higher neutron yields using \dt{}, due to the larger reaction cross-section for 100~keV deuterium ions; however, the 14~MeV neutrons produced by the \dt{} reaction are higher in energy than desired for low-energy nuclear recoil calibrations.
The \dd{} reaction provides neutrons with an average energy of 2.45~MeV, which is more appropriate for generating low-energy nuclear recoils with a measurable scattering angle in liquid noble targets. 
We will focus on the use of a \dd{} source in the following sections. 

The content is arranged as follows:
in Sec.~\ref{sec:dd_proposal} we propose a new neutron-scattering-angle-based nuclear recoil calibration technique for large liquid noble TPCs;
several potential enhancements to the newly proposed technique are described in Sec.~\ref{sec:advanced_dd_techniques}, including the creation of a monoenergetic 272~keV neutron source in Sec~\ref{sec:d_backscatter};
the neutron energy spectrum of a commercially available Adelphi Technology, Inc.\ DD108 neutron generator is measured in Sec.~\ref{sec:brown_dd_neutron_energy_spectrum_measurement} to demonstrate its suitability for the proposed nuclear recoil calibration techniques.

\section{Proposed nuclear recoil calibration using neutron scattering kinematics in a large liquid noble TPC} \label{sec:dd_proposal}

The current generation of liquid noble TPCs are commonly located at the center of large ($\mathcal{O}$(10~m) diameter) water tanks used to shield the TPC from unwanted external radioactive backgrounds during rare event searches~\cite{Agnes2015, AkeribAraujoBaiEtAl2015, AkeribAkerlofAkimovEtAl2015}.
A collimated beam of neutrons with known direction can be created by positioning a gas-filled (or evacuated) conduit inside the water tank spanning the space from the TPC cryostat to the wall of the water tank. 
A monoenergetic neutron source, such as a commercially available \dd{} neutron generator, placed outside the water tank in line with the conduit can be used to provide neutrons of fixed energy and direction into the TPC.
Using a 4m-long, 5~cm diameter neutron conduit with a neutron generator producing $10^{8}$~n/s into $4\pi$ solid angle, we expect $\sim$10$^{3}$~n/s incident upon the detector.
This incident neutron rate can be finely tuned by adjusting the duty cycle using available \dd{} generator pulsing capability.

This technique exploits the self-shielding properties of large TPCs to avoid contamination due to neutron scatters in passive materials that contribute to background events in more traditional \exsitu{} scattering-angle-based measurements.
Monte Carlo simulation studies of neutron transport in a realistic experimental setup indicate that the application of simple fiducial volume cuts in line with the neutron beam projection inside the TPC can ensure that 95\% of accepted events are produced by neutrons with energies within 6\% of the initial energy at the source~\cite{Malling2014}.
The collimated \dd{} neutron beam can also function as a very effective calibration source for the distribution of S2 vs.\ S1 for nuclear recoils.
The ratio of S2 to S1 is frequently used as a discriminant between nuclear and electronic recoils in liquid noble TPCs.
The neutron conduit can be aligned near the liquid noble target surface to provide a well-collimated beam of neutrons far from the reverse field region below the detector cathode---a common source of multiple scintillation, single ionization type event contamination in nuclear recoil band calibrations~\cite{Angle2007, LebedenkoAraujoBarnesEtAl2009}.

\begin{figure}[!htb]
    \begin{center}
        \includegraphics[width=0.480\textwidth]{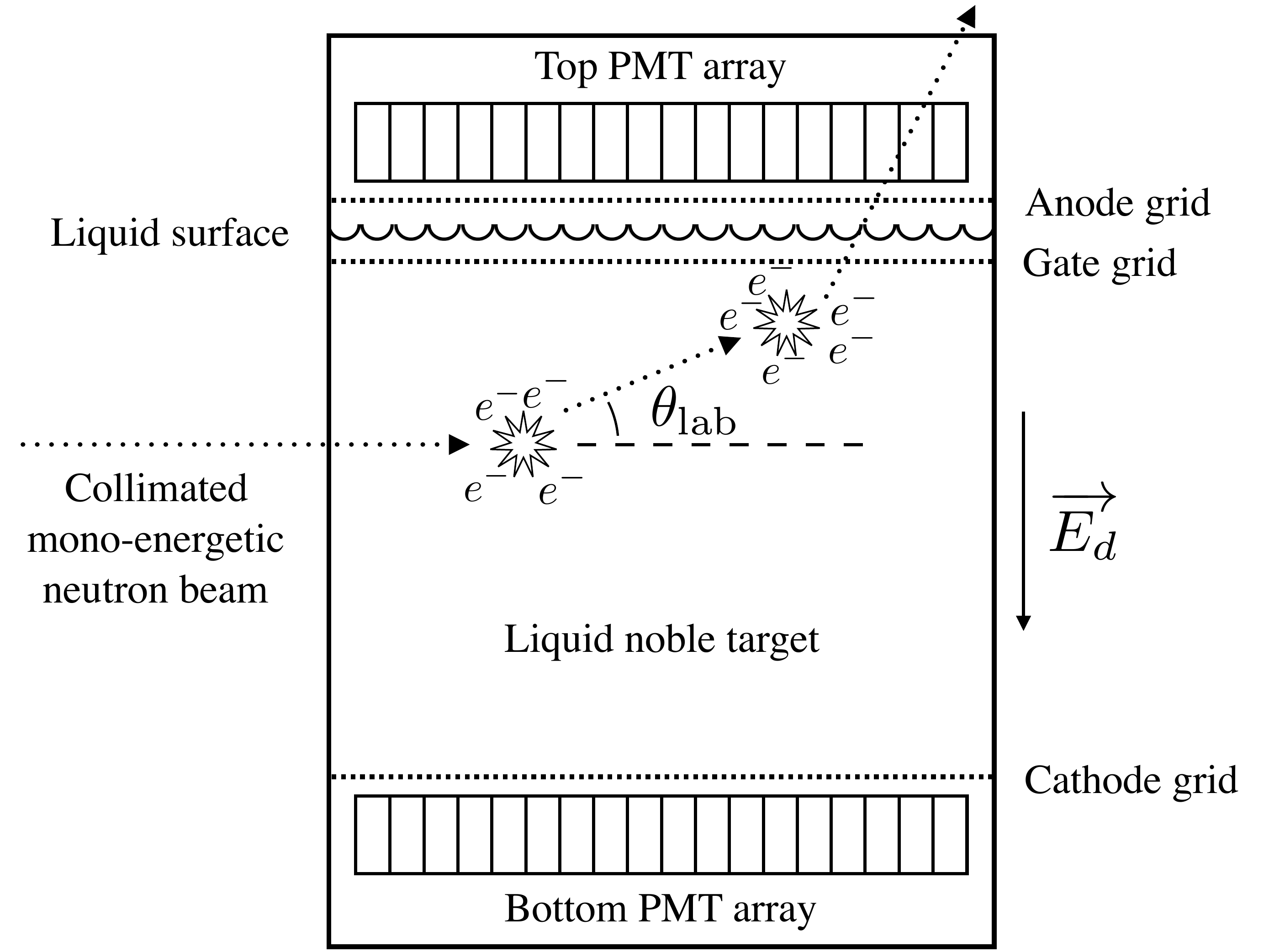}
        \vskip -0.1cm
        \caption{
            Diagram of a monoenergetic neutron scattering twice in a large TPC.
            The ($x$,\,$y$,\,$z$) position of both interactions can be reconstructed to provide a measurement of the scattering angle at the first vertex, $\theta_{\textrm{lab}}$.
            The prompt scintillation signals from each vertex typically overlap in the event record, but may be separately resolvable in some cases.
            The ionization signal from each vertex can be individually resolved in the event record for interactions separated in $z$ by a few mm.
            The signal generation and reconstruction parameters for liquid argon and xenon are listed in Table~\ref{tab:noble_liquid_tpc_parameters}.
            The nuclear recoil energy at the first scattering vertex can be reconstructed using the measured $\theta_{\textrm{lab}}$.
            The observed signals and measured energy at the first vertex provide a direct measurement of the signal yields.
        }
        \vskip -0.5cm
        \label{fig:conceptual_dd_scatter_diagram}
    \end{center}
\end{figure}

\medskip
\begin{table*}[htbp]
    \begin{threeparttable}
        \centering
        \caption{
            Relevant dual-phase liquid noble TPC parameters for liquid argon and xenon. 
        }   
        \label{tab:noble_liquid_tpc_parameters}
        \begin{tabular*}{\linewidth}{@{\extracolsep{\fill}} lcc}
            \toprule
            {Noble Target Characteristic} & {Ar \tnote{a}} & {Xe \tnote{b}} \\
            \midrule
            2.45~MeV total mean free path [cm] & 15 & 13 \\
            2.45~MeV elastic mean free path [cm] & 19 & 20 \\
            272~keV total mean free path [cm] & 18 & 14 \\
            272~keV elastic mean free path [cm] & 18 & 15 \\
            singlet lifetime [ns] & 6 \tnote{c} & 3.1 \tnote{d} \\
            triplet lifetime [ns] & $1.6 \times 10^{3}$ \tnote{c} & 24 \tnote{d} \\
            e$^{-}$ drift velocity in large TPCs [mm/$\mu$s] & $0.93 \pm 0.01$ (200~V/cm) \tnote{e} & $1.51 \pm 0.01$ (180~V/cm) \tnote{f} \\
            \bottomrule
        \end{tabular*}
        \begin{tablenotes}
        \item[a] Mean free paths calculated for $^{40}$Ar (99.6\% relative abundance) using Ref.~\cite{ShibataIwamotoNakagawaEtAl2011}.
        \item[b] Mean free paths calculated for natural Xe using Ref.~\cite{ShibataIwamotoNakagawaEtAl2011}.
        \item[c]~\cite{Hitachi1983}
        \item[d]~\cite{MockBarryKazkazEtAl2014}
        \item[e]~\cite{Agnes2015}
        \item[f]~\cite{Akerib2014}
        \end{tablenotes}
    \end{threeparttable}
\end{table*}
\medskip

The direct extraction of the signal yields depends upon the time structure of the S1 and S2 signals from each scattering vertex in the event record.
For scattering vertices separated by several mm in $z$, the S2 signal from each scattering vertex can be individually resolved in noble targets given the typically-achieved electron drift velocities of 1--2~mm/$\mu$s reported in Table~\ref{tab:noble_liquid_tpc_parameters}.
The ionization yield, \qy{}, of the target medium can be directly probed with an absolute measurement of nuclear recoil energy by fully reconstructing the scattering angle for multiple-vertex events and using the corresponding S2 information from each individual vertex.
Recent large TPCs using argon and xenon as the target media have achieved ($x$,\,$y$) position reconstruction uncertainties of $\mathcal{O}$(1~cm)~\cite{Brodsky2015, Faham2014} using the position reconstruction algorithm described in Ref.~\cite{SolovovBelovAkimovEtAl2012}.

The short timescale of the prompt S1 light makes direct extraction of the scintillation yield, \ly{}, more involved for some target media.
The 2.45~MeV neutrons produced by the \dd{} reaction have a velocity of 2.2~cm/ns.
In the case of liquid xenon, the similar singlet and triplet lifetimes in Table~\ref{tab:noble_liquid_tpc_parameters} combine to produce an S1 pulse envelope with a decay time of 20--30~ns.
The 45~ns time constant for electron-ion recombination in xenon is suppressed due to the drift field~\cite{Doke1999}.
Even the longest path lengths available in the current generation of liquid noble TPCs of $\sim$1~m provide a time separation between interactions that competes with the characteristic time constant of the S1 pulses themselves, leading to S1 pulse overlap in the event record.
Due to the large time difference between the singlet (6~ns) and triplet (1.6~$\mu$s) lifetimes for argon, the time structure of the prompt S1 light is dominated by photons produced by the singlet state; it may be possible to separate the singlet S1 contribution from each vertex in multiple-scatter events in that target media.

A direct, absolute calibration of \ly{} in multiple-scatter data using the observed neutron scattering angles can be achieved via a comparison of the S1 photon arrival times to the expected S1 pulse time structure given the location of the multiple neutron scattering vertices.
The measured S1 photon contribution from each vertex of known energy can be extracted via a maximum-likelihood-based comparison.
This pulse envelope time structure analysis promises to be more powerful when combined with the techniques described in Sec.~\ref{sec:advanced_dd_techniques}.
Alternatively, the scintillation yield can be extracted from the sample of single neutron scatters in line with the neutron beam projection in the TPC.
The absolutely calibrated S2 yield from the multiple-scatter \dd{} technique can be used to set the energy scale for observed single-scatter (1x S1, 1x S2) events, which allows for a precise extraction of the light yield via comparison with Monte Carlo simulation.

\section{Extension of the technique providing lower measured recoil energies and reduced calibration uncertainties} \label{sec:advanced_dd_techniques}

\subsection{Reduction of neutron bunch time structure} \label{sec:dd_short_pulse}

The neutron output of many commercially available \dd{} neutron generators can be pulsed.
The duration and frequency of the neutron pulses can be controlled using an external pulse generator.
The neutron generator model used in Sec.~\ref{sec:brown_dd_neutron_energy_spectrum_measurement} supports a nominal minimum pulse width of 100~$\mu$s.

Alternative pulsing solutions exist to provide neutron pulses with a duration as short as 1~ns to 1~$\mu$s~\cite{Csikai1987}.
Reducing the neutron bunch width time structure provides two powerful improvements over the technique described in Sec.~\ref{sec:dd_proposal}. 
First, narrowing the time envelope of the neutron pulse improves background rejection proportionally to the duty cycle.
Only events with prompt signals consistent with the generator pulse time and neutron propagation time to the detector are valid nuclear recoil candidates.
This allows rejection of backgrounds due to accidental coincidences and other spurious signals, which become increasingly prevalent when working close to threshold during the analysis of small nuclear recoil signals.
This delivers a lower energy threshold and reduced systematics for the calibration signal event set.
Second, when the neutron pulse width becomes sufficiently narrow, it can be used to establish the $t_{0}$ for the electron drift of a scattering event.
This $t_{0}$ information is traditionally provided by the S1 in dual-phase TPCs. 
Establishing the $t_{0}$ independently of the observation of an S1 signal permits the investigation of the S2 associated with neutron scatter events that are so low in recoil energy that the associated S1 signal is typically undetected.
In liquid xenon, we expect this to potentially extend the S2 signal yield studies down to $\mathcal{O}$(100~\evnr{}).
This ultra-low-energy charge yield calibration technique is significant for determining the sensitivity of TPC experiments to low-mass WIMPs using S2-only searches.

The $z$ position of a particle interaction in these detectors is typically determined by measuring the electron drift time using the pulse timing information provided by the S1 and S2 signals.
The known electron drift velocity (Table~\ref{tab:noble_liquid_tpc_parameters}) for a given electric drift field $\vec{E_{d}}$ applied across the target media allows the reconstruction of the $z$ position with a precision of ${\sim} 1$~mm.
A reduction in the neutron bunch width time structure to 10~$\mu$s will provide position reconstruction of S2-only events in the $z$ dimension with a resolution of roughly 2~cm, similar to the ($x$,\,$y$) reconstruction precision provided by PMT top array hit-pattern analysis of S2 signals.
The ability to reconstruct S2-only events to high precision in three dimensions allows for the identification of candidate S2-only events that are consistent with a neutron interaction in the detector given the expected drift time for events in line with the beam pipe.
It is then possible to determine the number of single-scatter events with zero detected photons for a given observed S2 pulse size.
This allows for an additional \ly{} calibration technique providing stronger statistical constraints on the S1 yield. 

Further narrowing the neutron bunch to a width of \bigo{(100~ns)} or shorter may be possible.
This improved time definition of the neutron pulse would permit the use of time-of-flight (ToF) energy tagging for neutrons generated by the \dd{} source.
The neutrons of interest for this type of calibration scatter in a deliberately positioned hydrogenous moderator outside of the water tank near the neutron generator, yielding a sample of neutrons with a broad spectrum of kinetic energies traveling down the beam pipe to the TPC.
The measured ToF would then provide the neutron kinetic energy on a per-event basis.
The calculated ToF for neutrons from 1--2450~keV is shown in Table~\ref{tab:neutron_tof_energy_dependence}. 
Assuming a 4~m beam path from the hydrogenous moderator to the TPC active region, moderated neutrons ranging from 1--2450~keV would have an expected ToF between 200~ns and 10~$\mu$s.
It may be possible to use a fast organic scintillator such as BC501A as the hydrogenous moderator, and establish the neutron ToF $t_{0}$ using the organic scintillator pulse. 

\medskip
\begin{table}[htbp]
    \centering
    \caption{
        The time-of-flight (ToF) dependence upon neutron energy.
        The corresponding nuclear recoil spectrum endpoint energy in argon and xenon is given in columns three and four, respectively.
    }
    \label{tab:neutron_tof_energy_dependence}
    \begin{tabular*}{\columnwidth}{@{\extracolsep{\fill}} SSSS}
        \toprule
        {$E_{n}$ [keV]} & {ToF [ns/m]} & \multicolumn{2}{c}{Maximum Recoil [\kevnr{}]} \\
        \cmidrule{3-4}
        &  & {Ar} & {Xe} \\
        \midrule
        1 & 2286 & 0.1 & 0.03 \\
        10 & 723 & 1 & 0.3 \\
        100 & 229 & 10 & 3 \\
        272 & 139 & 26 & 8 \\
        1000 & 72 & 96 & 30 \\
        2450 & 46 & 235 & 74 \\
        \bottomrule
    \end{tabular*}
\end{table}
\medskip

\subsection{Reduction in neutron energy using a deuterium-loaded reflector} \label{sec:d_backscatter}

The technique described in this section can provide an inexpensive and portable quasi-monoenergetic source 272~keV neutrons that can be used to extend the kinematic calibration (described in Sec.~\ref{sec:dd_proposal}) nearly an order of magnitude lower in energy.
This lower-energy source is well matched to the nuclear recoil energy region used for low-mass WIMP searches and the expected coherent elastic neutrino-nucleus scattering (CENNS) signal in upcoming large liquid noble dark matter detectors~\cite{Strigari2009, BillardFigueroa-FelicianoStrigari2014}. 

\subsubsection{A monoenergetic 272~keV neutron source}

A beam of quasi-monoenergetic 272~keV neutrons can be obtained by positioning a deuterium-loaded material (the ``reflector'') behind the \dd{} neutron generator, directly in line with the neutron collimation conduit leading to the TPC (see Fig.~\ref{fig:deuterium_backscatter_reflector_sim_geometry}).
The limited solid angle presented by the neutron conduit is used to collect neutrons that scatter in the deuterium-loaded reflector with a scattering angle of $\sim$180$^{\circ}$.
Deuterium is an optimal reflector material; its low atomic mass provides the most significant reduction in neutron energy possible for $\sim$180$^{\circ}$ elastic scatters~\cite{Gaitskell2015}---larger energy reductions from neutron scatters on $^{1}$H are discussed addressed below.
These reflected neutrons have a minimum kinetic energy of 272~keV.
In addition, a double-scatter (both scatters must be neutron-deuteron) elastic scattering event with a summed scattering angle of 180$^{\circ}$ within the deuterium-loaded reflector also provides an outgoing 272~keV neutron.

Although neutron-hydrogen scattering can result in neutron energies below 272~keV, all neutron scatters when using hydrogen are in the forward direction with a scattering angle of 0--90$^{\circ}$ in the lab frame.
With a hydrogen reflector, small variations in the neutron scattering angle produce large fluctuations in reflected neutron energy. 
In contrast, using direct backscatters provided by deuterium's significant differential scattering cross-section at 180$^{\circ}$ suppresses the effects of variations in scattering angle, and provides a better defined quasi-monoenergetic neutron beam.
Deuterium has the largest cross-section for 180$^{\circ}$ scatters of all potential reflector materials.

The $\times 9$ reduction in the neutron beam energy provided by the deuterium reflector has several advantages for low-energy nuclear recoil calibration.
The use of 272~keV neutrons provides a reduction in the uncertainty associated with kinematic energy reconstruction for low-energy events.
A 1~\kevnr{} nuclear recoil produced by a 2.45~MeV neutron in liquid xenon corresponds to a neutron scattering angle of 13$^{\circ}$, which is a 4.6~cm deflection over a length of 20~cm.
By comparison, a 1~\kevnr{} nuclear recoil produced by a 272~keV neutron in liquid xenon has a scattering angle of 41$^{\circ}$, which is a 14~cm deflection over the same vertex separation. 
In large liquid xenon TPCs, the typical uncertainty associated with ($x$,\,$y$) position reconstruction of each vertex in events of this nuclear recoil is 1--3~cm~\cite{Verbus2016}. 
We estimate that in the 1--4~\kevnr{} range, the ($\sigma/\mu$) resolution for angle-based recoil energy reconstruction may be improved by a factor of $\times 2$ due to this increase in the average scattering angle for a given recoil energy.
The increased scattering angle for nuclear recoils of a given energy improves the efficiency of the detection of calibration events below 1~\kevnr{}. 
This improved efficiency allows the technique to directly measure recoil energies of $\mathcal{O}$(100~\evnr{}) in liquid xenon, where the expectation is $\sim$1 ionization electrons at 180~V/cm~\cite{Verbus2016}.

This neutron reflection technique reduces the neutron flux incident on the TPC by $\times 1/450$ compared to the direct \dd{} source calibration described in Sec.~\ref{sec:dd_proposal} when using similar neutron generator operating parameters and experimental geometry.
The reduction of the relative event rate in the TPC can be more than compensated for by the use of the following techniques:

\begin{enumerate}[i.]
    \item Increase the \dd{} source neutron flux from $10^{7}$~n/s to $10^{9}$~n/s.
    \item Expand the neutron conduit diameter from 5~cm to 15~cm.
        For a typical experimental configuration, this larger neutron conduit diameter increases the angular acceptance from ${\pm} 0.4^{\circ}$ to ${\pm} 1^{\circ}$, which provides $\times 9$ greater neutron flux.
        The diameter of the reflector can also be correspondingly increased by a factor of $\times 3$.
        The combination of these effects provides a $\times 60$ increase in the neutron flux entering the TPC for the dimensions shown in Fig.~\ref{fig:deuterium_backscatter_reflector_sim_geometry}.
    \item The larger differential scattering cross-section in xenon for 272~keV neutrons compared to 2.45~MeV neutrons provides a $\times2.5$ increase in the low-energy recoil rate.
\end{enumerate}

\subsubsection{Simulation of the deuterium-loaded neutron reflector}

\begin{figure*}[!htb]
    \begin{center}
        \includegraphics[width=0.95\textwidth]{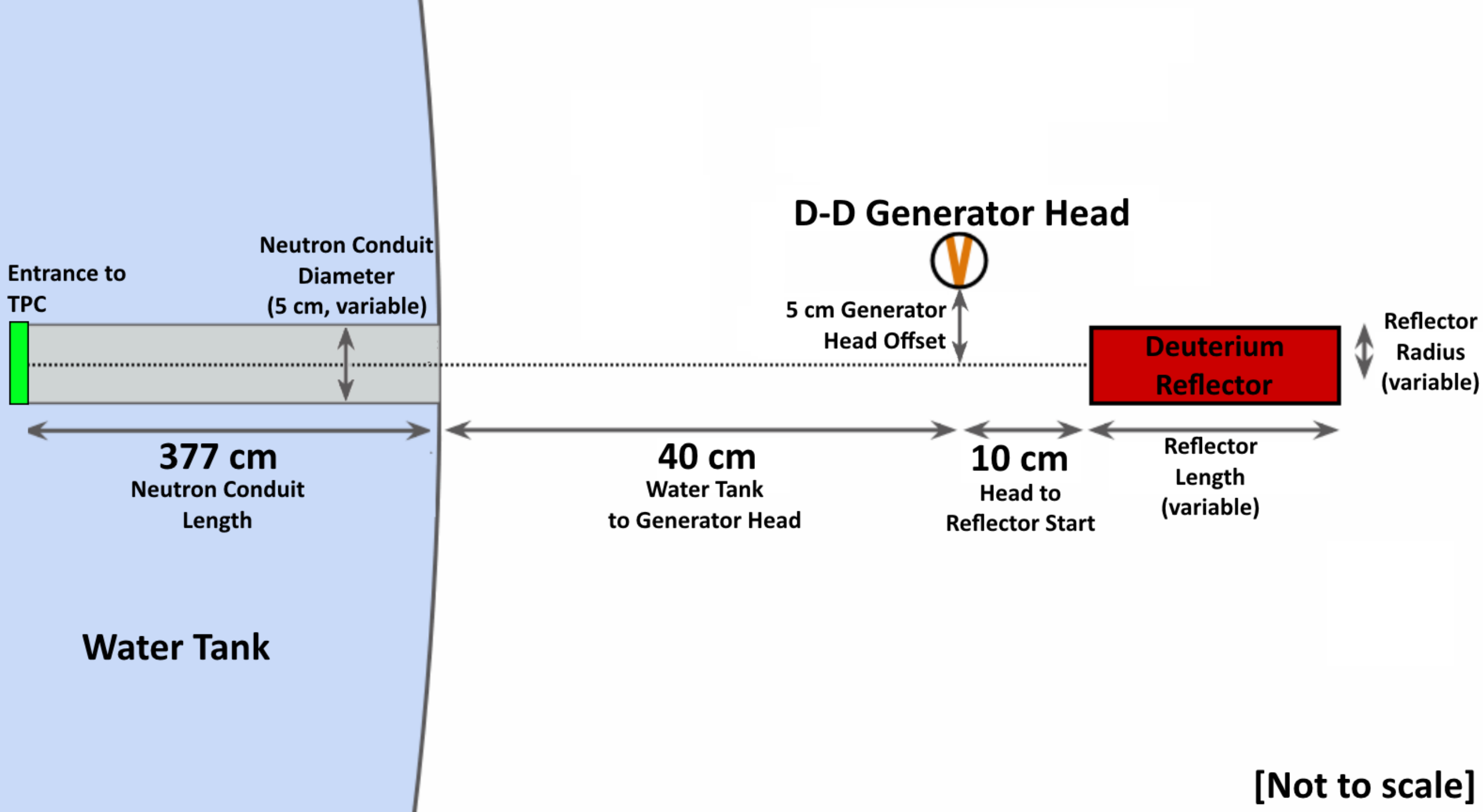}
        \vskip -0.1cm
        \caption{
            Simulation geometry setup for neutron reflector studies. 
            Neutrons produced by the \dd{} source elastically scatter through an angle $\sim$180$^{\circ}$ in the deuterium reflector and are selected by the solid angle of the neutron conduit.
            The reflector material type, length, and radius as well as the generator head offset were individually varied to determine the optimal configuration.
            The neutron generator must be placed out of line with the neutron conduit to eliminate line-of-sight 2.45 MeV neutrons from the \dd{} source entering the TPC.
        }
        \vskip -0.5cm
        \label{fig:deuterium_backscatter_reflector_sim_geometry}
    \end{center}
\end{figure*}

In order to optimize the technique and select the best type of deuterium-based reflector, a series of Geant4-based~\cite{AgostinelliAllisonAmakoEtAl2003} (Geant4 version 4.9.4.p04 using G4NDL3.14) simulations using the geometry shown in Fig.~\ref{fig:deuterium_backscatter_reflector_sim_geometry} were performed.
To eliminate contamination at the TPC from 2.45~MeV neutrons, the \dd{} neutron generator must have a non-zero offset from the center of the calibration conduit leading to the TPC. 
Given the dimensions in Fig.~\ref{fig:deuterium_backscatter_reflector_sim_geometry} and a generator head offset of 5~cm from the beam line, reflected neutrons collected by the neutron conduit have a mean energy of 290~keV. 
The neutron generator offset, reflector length, reflector size, and reflector type were varied to study the effects on the resultant neutron energy spectrum at the TPC.
Each simulated reflector configuration was evaluated based on both the number of reflected neutrons reaching the TPC within a usable energy band and the beam contamination from neutrons of other energies. 
The figure of merit for the energy contamination study was the ratio of the number of neutrons entering the TPC with energies within $\pm$10\% of the reflected neutron peak to the number with energies $<$1~MeV, henceforth referred to as ``beam purity.''
Nuclear recoils from neutrons entering the TPC with energies $>$1~MeV can be rejected in analysis based upon the size of the ionization signal relative to the measured scattering angle. 

Three potential reflector types were considered in this study: gaseous D$_{2}$, liquid D$_{2}$, and heavy water (D$_{2}$O).
The gaseous D$_{2}$ reflector was simulated with a gas pressure achievable in existing cylinders (340~bar, density of 0.047~g/cm$^{3}$) and with the surrounding container materials from an available commercial product (the Luxfer T45J, a carbon-fiber-based cylinder).
The impact of varying container thicknesses for a gaseous D$_{2}$ reflector was studied. 
The liquid D$_{2}$ reflector (density of 0.16~g/cm$^{3}$) was simulated without containment to demonstrate the highest achievable performance. 
Simulations of pure (gaseous or liquid) D$_{2}$ reflectors were used to independently vary aspects of the geometry (such as reflector radius, length, orientation, end cap shape, density, generator offset and conduit radius) to determine the impact of each parameter on the resulting neutron spectrum at the TPC.
The heavy water reflector (D$_{2}$O) was simulated without containment; its container can be negligibly thin-walled in practice.
The effects due to the oxygen atoms in the D$_{2}$O were studied. 

\begin{figure}[!htb]
    \begin{center}
        \includegraphics[width=0.48\textwidth]{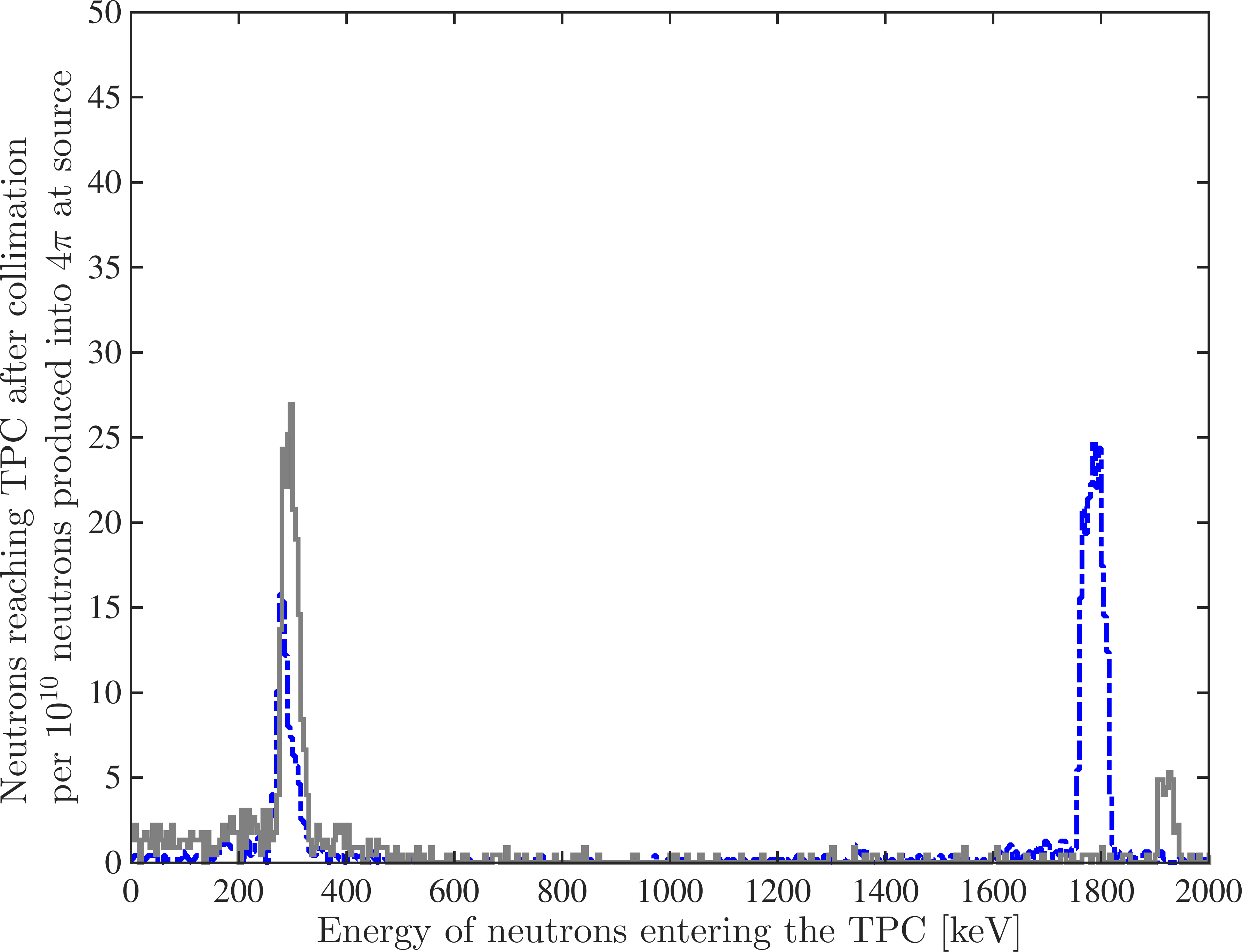}
        \vskip -0.1cm
        \caption{
            Gaseous D$_{2}$ cylinder vs.\ D$_{2}$O reflected neutron spectrum comparison.
            The simulated energy spectra of neutrons incident upon the TPC are shown after scattering in either the gaseous D$_{2}$ (blue dashed-dotted) or D$_{2}$O (gray solid) reflectors. 
            The gaseous D$_{2}$ reflector used a container geometry based upon the Luxfer T45J carbon fiber cylinder.
        }
        \vskip -0.5cm
        \label{fig:gD2_vs_D2O_comparison}
    \end{center}
\end{figure}

Representative simulation results comparing gaseous D$_{2}$ and D$_{2}$O reflector media with a 5~cm offset neutron generator head are shown in Fig.~\ref{fig:gD2_vs_D2O_comparison}.
The gaseous D$_{2}$ and D$_{2}$O reflectors were set to have an identical orientation and position for both simulation trials. 
The low-energy neutron peak produced by both types of reflector media is visible at $\sim$300~keV.
The two peaks observed at higher energy in the gaseous D$_{2}$ and D$_{2}$O simulations are produced by neutrons that interact with passive container materials and neutron-oxygen scatters in the reflector, respectively. 
The energy purity figure of merit is nearly identical for gaseous D$_{2}$ and D$_{2}$O (57\% purity in D$_{2}$ compared to 60\% in D$_{2}$O); however, the use of the D$_{2}$O results in a $\times 2.3$ increase in the reflected neutron flux over gaseous D$_{2}$ within $\pm$10\% of the reflected neutron peak after collimation.
D$_{2}$O is a more favorable reflector material based upon the defined energy purity and neutron flux criteria when compared to a gaseous D$_{2}$ reflector at the pressures achievable using thin-walled commercially available containment (Luxfer's T45J cylinder). 

\begin{figure}[!htb]
    \begin{center}
        \includegraphics[width=0.48\textwidth]{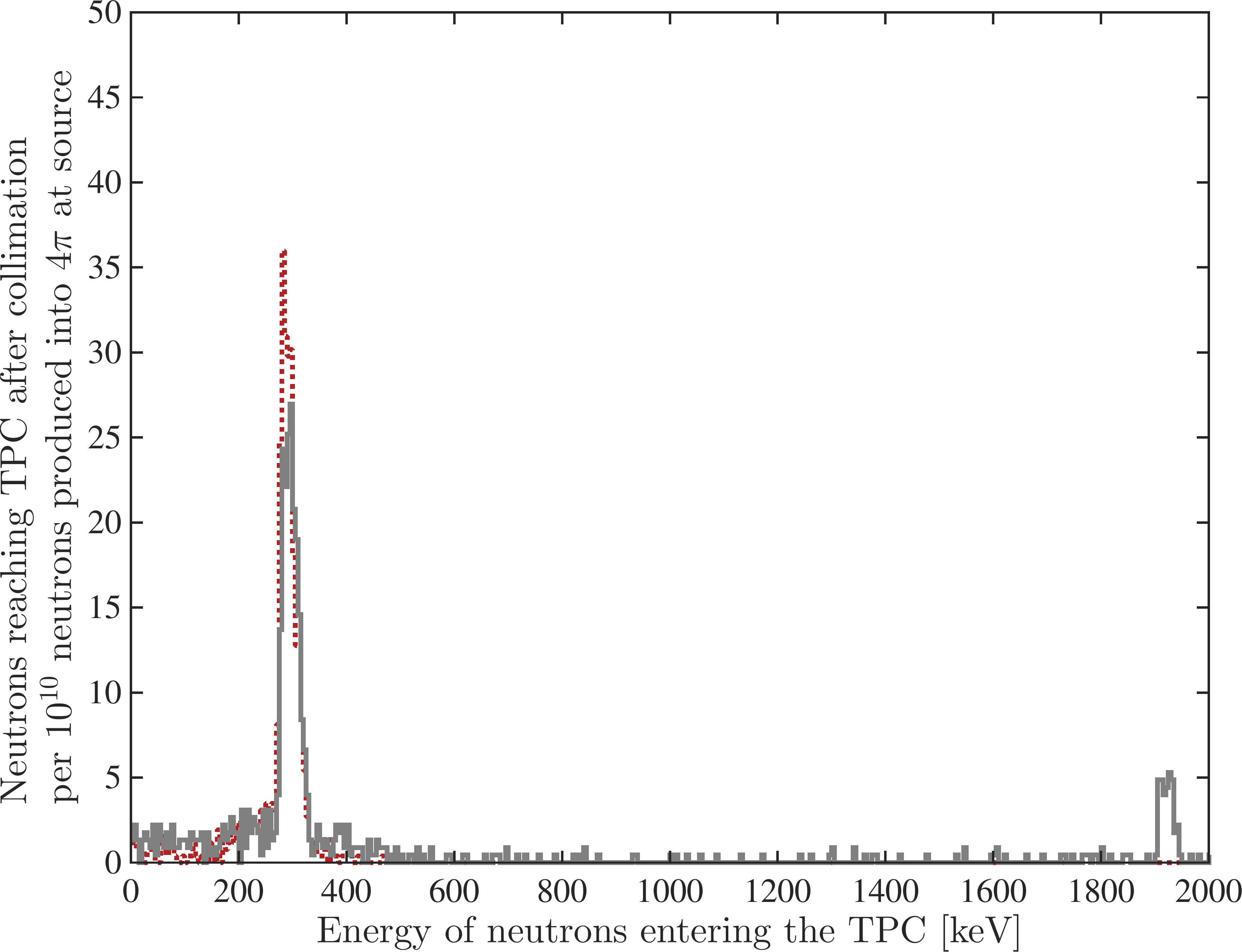}
        \vskip -0.1cm
        \caption{
            Liquid D$_{2}$ vs.\ D$_{2}$O reflected neutron spectrum comparison.
            The simulated energy spectra of neutrons incident upon the TPC are shown after scattering in either the liquid D$_{2}$ (red dotted) or D$_{2}$O (gray solid) reflectors. 
        }
        \vskip -0.5cm
        \label{fig:D2O_vs_LD2_comparison}
    \end{center}
\end{figure}

Representative simulation results comparing the D$_{2}$O and liquid D$_{2}$ reflector media are shown in Fig.~\ref{fig:D2O_vs_LD2_comparison}. 
The liquid D$_{2}$ reflector modestly outperformed both the gaseous D$_{2}$ and D$_{2}$O reflectors in terms of energy purity (67\%).
The liquid D$_{2}$ also provided the largest low-energy neutron flux incident upon the TPC.
In addition, the liquid D$_{2}$ can better scale to larger reflector sizes---with correspondingly larger low-energy quasi-monoenergetic neutron fluxes incident on the TPC---than the D$_{2}$O reflector for which the useful size is limited due to off-energy neutron contamination from neutron interactions with oxygen atoms in the reflector media.
These results indicate that a liquid D$_{2}$ reflector could potentially exceed the performance of a D$_{2}$O reflector; however, pure D$_{2}$ targets require a significantly more complex experimental setup.

\subsubsection{TPC calibration backgrounds when using the deuterium neutron reflector}

Neutrons interacting with materials other than deuterium in the reflector setup provide a source of high-energy neutron contamination in the neutron energy spectrum at the TPC. 
Secondary oxygen recoils in the D$_{2}$O reflector create a high-energy neutron background that scales with the reflector mass. 
A size restriction on effective D$_{2}$O based reflector is set by the neutron mean free path between oxygen recoils in the reflector media.
Increasing both the D$_{2}$O reflector diameter and the neutron conduit diameter from 5~cm to 15~cm results in a $\times 60$ increase in flux.
This increase in flux comes at the cost of neutron energy purity; there is more than a $\times 2.1$ drop in the beam purity due to oxygen recoils in the reflector. 
For comparison, a liquid D$_{2}$ reflector enlarged in the same way results in an equivalent proportional flux increase, but only a $\times 1.6$ reduction in beam purity.
It was found that the performance of the gaseous D$_{2}$ reflector can also be improved by increasing the D$_{2}$ density and by reducing the containment wall thicknesses.
While a D$_{2}$O reflector is currently the most effective and easily deployable option, the scaling properties and lack of oxygen in the reflector media make pure D$_{2}$ reflectors a compelling topic for further study. 

A simulation of the water shielding surrounding the neutron conduit was used to estimate the relative magnitude of contaminating background effects due to neutrons scattering in the water.
For a \dd{} source with a 5~cm offset from the neutron conduit, 13\% of the neutrons entering the TPC have lost energy in the water shield.
The simulation indicates that 90\% of the neutrons entering the TPC after losing energy in the water are either substantially above ($>$1000~keV) or substantially below ($<$1~keV) the energy region of interest for reflected neutrons and would not interfere with TPC calibration.
The result is that 98\% of neutrons entering the TPC in the energy range 1--1000~keV are direct neutrons from the deuterium reflector. 
Additional possible contamination from neutrons that have scattered in the D-D source hardware can be suppressed by ensuring the neutron generator assembly is sufficiently out of line with the neutron conduit. 

\begin{figure}[!htb]
    \begin{center}
        \includegraphics[width=0.48\textwidth]{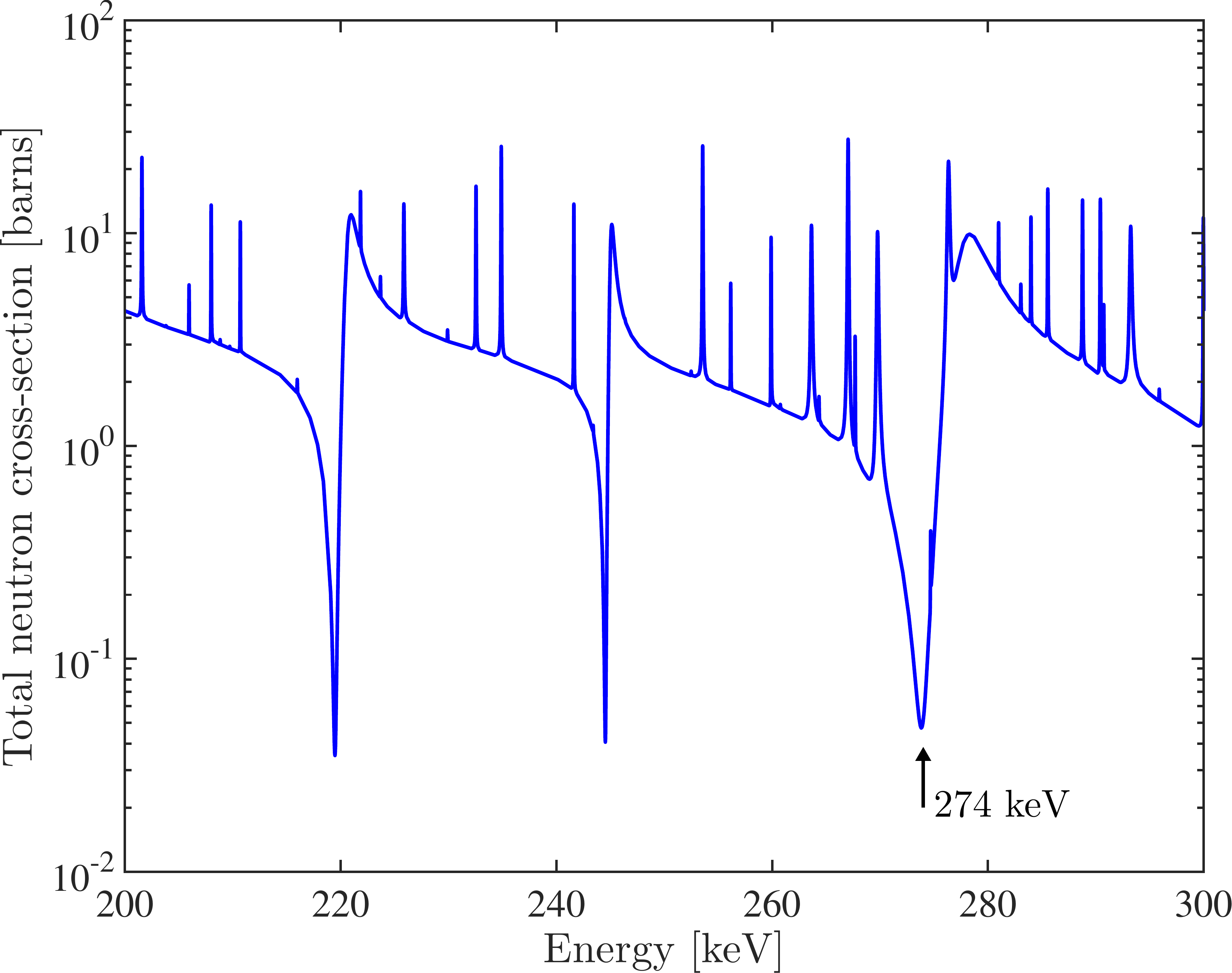}
        \vskip -0.1cm
        \caption{
            The total neutron cross-section for $^{56}$Fe evaluated using the ENDF/B-VII.1 nuclear database~\cite{ChadwickHermanOblozinskyEtAl2011}.
        }
        \vskip -0.5cm
        \label{fig:fe56_total_neutron_cross_section}
    \end{center}
\end{figure}

A notch in the $^{56}$Fe neutron scattering cross-section at 274 keV suggests a method for improving the energy distribution of reflected neutrons.
The notch is shown in Fig.~\ref{fig:fe56_total_neutron_cross_section}.
A similar neutron energy filter technique was used in Ref.~\cite{Joshi2014}. 
By placing a 2.5~cm radius iron cylinder in line between the generator and the neutron conduit, all neutrons except those at the desired low-energy peak can be eliminated at the cost of neutron beam intensity. 
This effect can be used to reduce contamination from off-energy neutrons and improve the width of the energy distribution of neutrons entering the TPC.

\subsection{Direct scintillation yield measurement using the S1 photon arrival time structure} \label{sec:direct_s1_yield_photon_time_structure}

Analysis of the S1 pulse envelope time structure can provide a direct \ly{} calibration using the measured scattering angle between neutron interactions. 
For double-scatter events in the TPC with a given vertex separation, the time separation between the S1 signals from each scatter in the event record is ${\propto} 1 / \sqrt{E_{n}}$ as determined by the neutron travel time between interactions in the target media. 
The ToF for a variety of neutron energies is shown in Table~\ref{tab:neutron_tof_energy_dependence}. 
We will use a 50~cm path between neutron interactions as a benchmark value---a length comfortably contained within the liquid noble target of upcoming TPCs~\cite{AkeribAkerlofAkimovEtAl2015}.
For the direct 2.45~MeV neutrons from the \dd{} source, a double-scatter event with 50~cm of separation between neutron interactions in the TPC has a 23~ns difference between the time of the first and second scatters.
After scattering once, the probability of a 2.45~MeV neutron traveling $\geq 50$~cm before scattering again in the target media is 2\% for xenon and 4\% for argon, based upon the total mean free path values in Table~\ref{tab:noble_liquid_tpc_parameters}.
Given the time constants for S1 signal generation in Table~\ref{tab:noble_liquid_tpc_parameters}, it is possible to perform a likelihood-based analysis on the pulse shape envelope of the scintillation signal to determine the photon contribution from the first scatter. 
As mentioned in Sec.~\ref{sec:dd_proposal}, argon scintillation lends itself particularly well to this technique due to the short lifetime of the singlet S1 component. 

A more clear identification of the S1 contributions from the first and second scatters is made possible when using 272~keV neutrons produced by the reflector source described in Sec.~\ref{sec:d_backscatter}. 
The corresponding time difference over the same 50~cm path length between scatters when using 272~keV neutrons is 70~ns. 
This $\times 3$ reduction in neutron velocity compared to the direct 2.45~MeV neutrons from the \dd{} source provides sufficient time separation to clearly identify the S1 photons from each individual interaction in a liquid noble TPC.
After scattering once, the probability of a 272~keV neutron traveling $\geq 50$~cm before scattering again in the target media is 3\% for xenon and 5\% for argon, based upon the total mean free path values in Table~\ref{tab:noble_liquid_tpc_parameters}.
The scattering-angle-based measurement of the recoil energy at the first scattering vertex can be compared to the observed number of S1 photons from that interaction to provide a direct measurement of the \ly{} similar to the \qy{} measurement described in Sec.~\ref{sec:dd_proposal}.

\section{D-D neutron generator energy spectrum measurement} \label{sec:brown_dd_neutron_energy_spectrum_measurement}

A monoenergetic source of neutrons is required to perform the absolute calibrations described in Sec.~\ref{sec:dd_proposal} and Sec.~\ref{sec:advanced_dd_techniques}.
Commercially available \dd{} neutron generators meet all of the criteria outlined in Sec.~\ref{sec:introduction}.
In this section, we characterize the neutron energy spectrum produced by an Adelphi Technology, Inc.\ DD108 neutron generator using neutron ToF over a known distance to determine its suitability as a neutron source for the proposed kinematics-based TPC calibrations~\cite{Vainionpaa2013, Vainionpaa2014}.

\subsection{Adelphi Technology, Inc.\ DD108 neutron generator}

\begin{figure}[!htb]
    \begin{center}
        \includegraphics[width=0.480\textwidth]{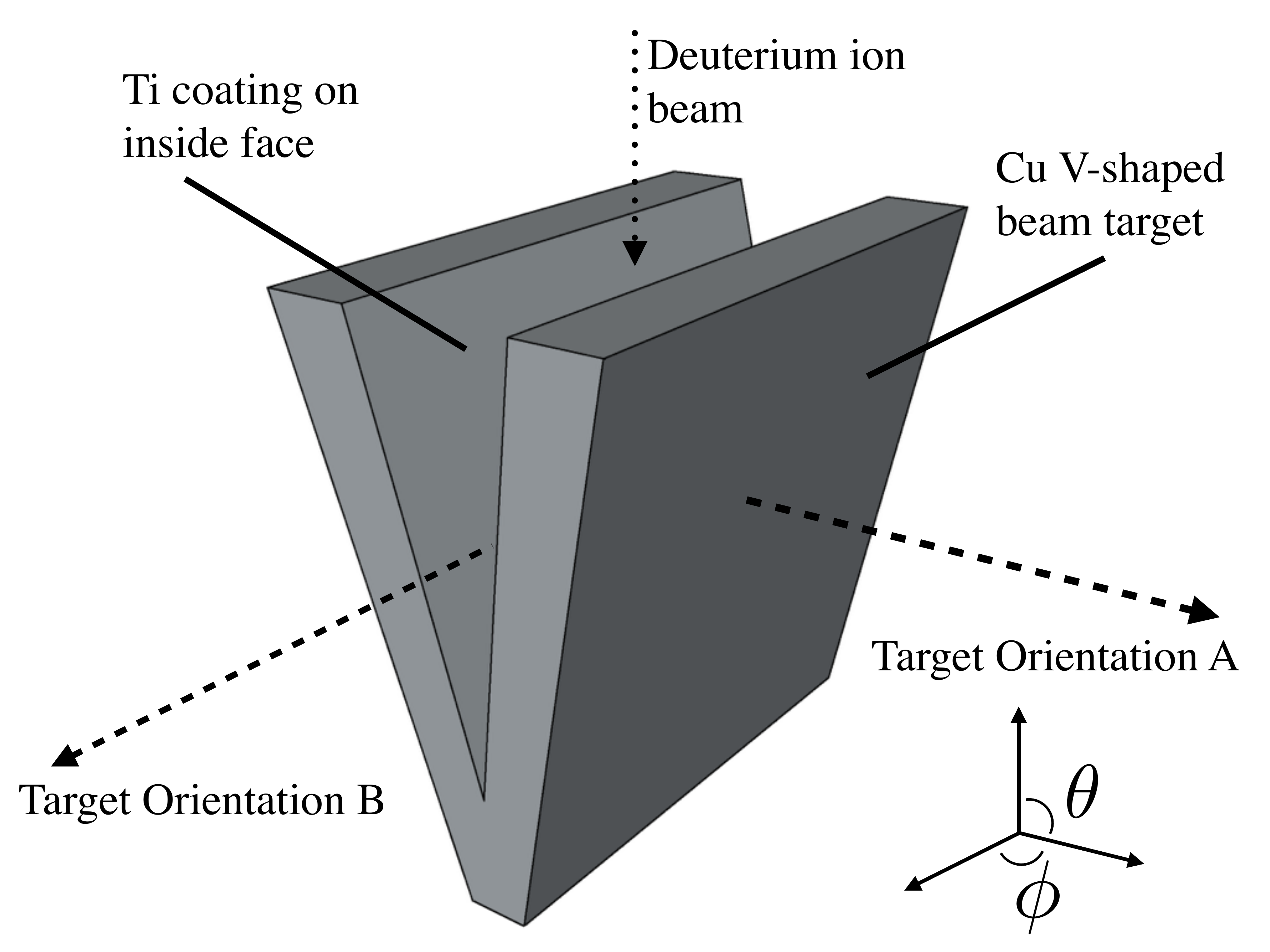}
        \vskip -0.1cm
        \caption{
            A conceptual diagram of the copper V-shaped beam target of the DD108 neutron generator.
            The deuterium ion beam is incident upon the target from the top of the figure.
            The ToF energy spectrum measurements were made at $\theta = \pi/2$.
            The ToF measurement at Target Orientation A ($\phi = 0$) was off-axis with the target V, and Target Orientation B ($\phi = -\pi/2$) was on-axis with the target V.
            This $\pi/2$ variation in $\phi$ between the off-axis and on-axis measurements was used to quantify any variation in the neutron spectrum due to the asymmetry of the neutron production surface.
        }
        \vskip -0.5cm
        \label{fig:dd108_beam_target.pdf}
    \end{center}
\end{figure}

\begin{figure}[!htb]
    \begin{center}
        \includegraphics[width=0.480\textwidth]{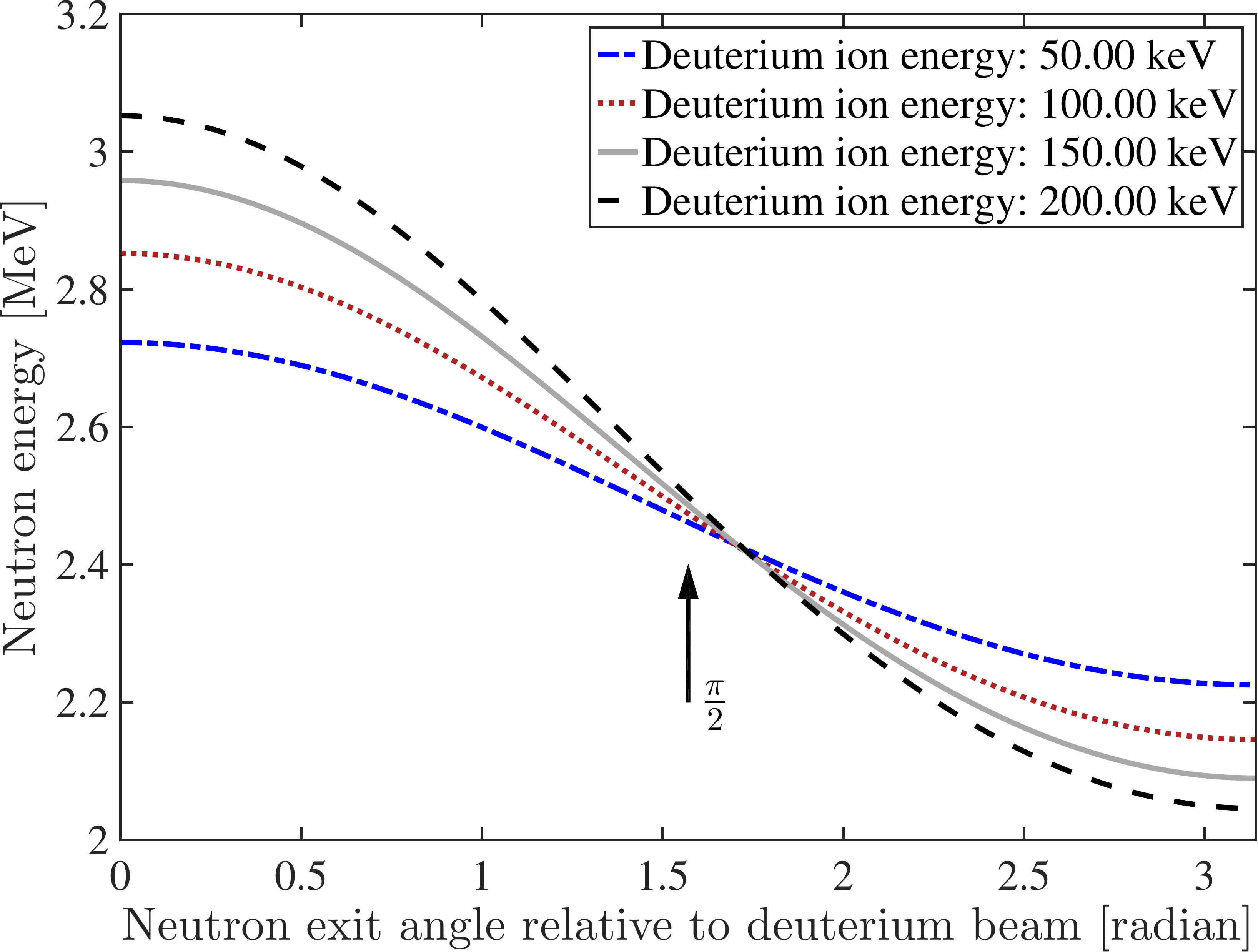}
        \vskip -0.1cm
        \caption{
            The neutron energy produced by the $^2$H(d,\,n)$^3$He fusion reaction as a function of angle.
            We show this function for several incident deuterium ion energies.
            The dependence of the neutron energy on the acceleration potential can be reduced by using neutrons produced at $\pi/2$ relative to the deuterium ion beam.
            Figure produced using information from Ref.~\cite{Csikai1987}.
        }
        \vskip -0.5cm
        \label{fig:dd_neutron_energy_vs_angle}
    \end{center}
\end{figure}

The DD108 is a beam-on-target \dd{} neutron generator with a nominal maximum neutron output of $10^8$~n/s.
Inside the DD108, deuterium ions are accelerated across a $\sim$100~kV potential difference into a titanium-coated copper target as shown in Fig.~\ref{fig:dd108_beam_target.pdf}.
The incident deuterium ions chemically bond with the titanium coating.
Subsequent incident deuterium ions fuse with the captured deuterium in the target and produce neutrons into 4$\pi$ solid angle via the $^{2}$H(d,\,n)He$^{3}$ reaction.
The mean outgoing neutron energy and flux are functions of the incident deuterium ion energy.
The neutron energy and flux are also dependent upon the angle between the deuterium ion beam and the outgoing neutron direction~\cite{Csikai1987}.
The neutron flux varies by about a factor of $\times 2$ as a function of polar angle relative to the \dd{} generator beam target.
The neutron energy as a function of angle relative to the deuterium ion beam is shown in Fig.~\ref{fig:dd_neutron_energy_vs_angle} for a range of acceleration potentials.

The physical size and shape of the deuterium ion beam target can have an effect on the mean energy and width ($\sigma$/$\mu$) of the neutron spectrum produced due in part to deuterium ion straggling in the target material before neutron production~\cite{Csikai1987}.
The DD108 target is V-shaped, as shown in Fig.~\ref{fig:dd108_beam_target.pdf}, to present an increased surface area to the incident deuterium ion beam for heat dissipation purposes.
The dependence of the observed neutron spectrum upon the kinetic energy of the accelerated deuterium ion can be reduced by using outgoing neutrons at $\theta = \pi/2$ relative to the generator ion beam for nuclear recoil calibration purposes.

The neutron spectrum was measured using two separate DD108 target orientations to quantify any effects due to target asymmetry, and determine if there is an optimal configuration for the nuclear recoil calibrations described in Sec.~\ref{sec:dd_proposal} and Sec.~\ref{sec:advanced_dd_techniques}.
Target Orientation A ($\theta = \pi/2$, $\phi = 0$) measured the ToF spectrum of neutrons escaping perpendicular to the axis of the V-target, and Target Orientation B ($\theta = \pi/2$, $\phi = -\pi/2$) did the same for neutrons escaping parallel to the axis of the V-target. 

The only gamma rays produced in the generator are via the $^2$H(d,\,$\gamma$)$^4$He reaction with an energy of 23.84~MeV, which is suppressed by a factor of ${\sim} 10^{-7}$ relative to the neutron production rate~\cite{Wilkinson1985}.
This corresponds to roughly 10 $\gamma$/s when operating at the nominal maximum DD108 neutron yield of $10^{8}$~n/s. 
Electrons liberated by ion impacts on the target can back-stream across the 125~kV potential in the neutron generator and produce bremsstrahlung x-rays upon interaction with structural materials~\cite{IAEA2012}.
The V-shaped beam target is surrounded by a shroud electrode biased to a slightly higher voltage in order to prevent x-ray production or hardware damage by collecting the back-streaming electrons. 

\begin{figure}[!htb]
    \begin{center}
        \includegraphics[width=0.480\textwidth]{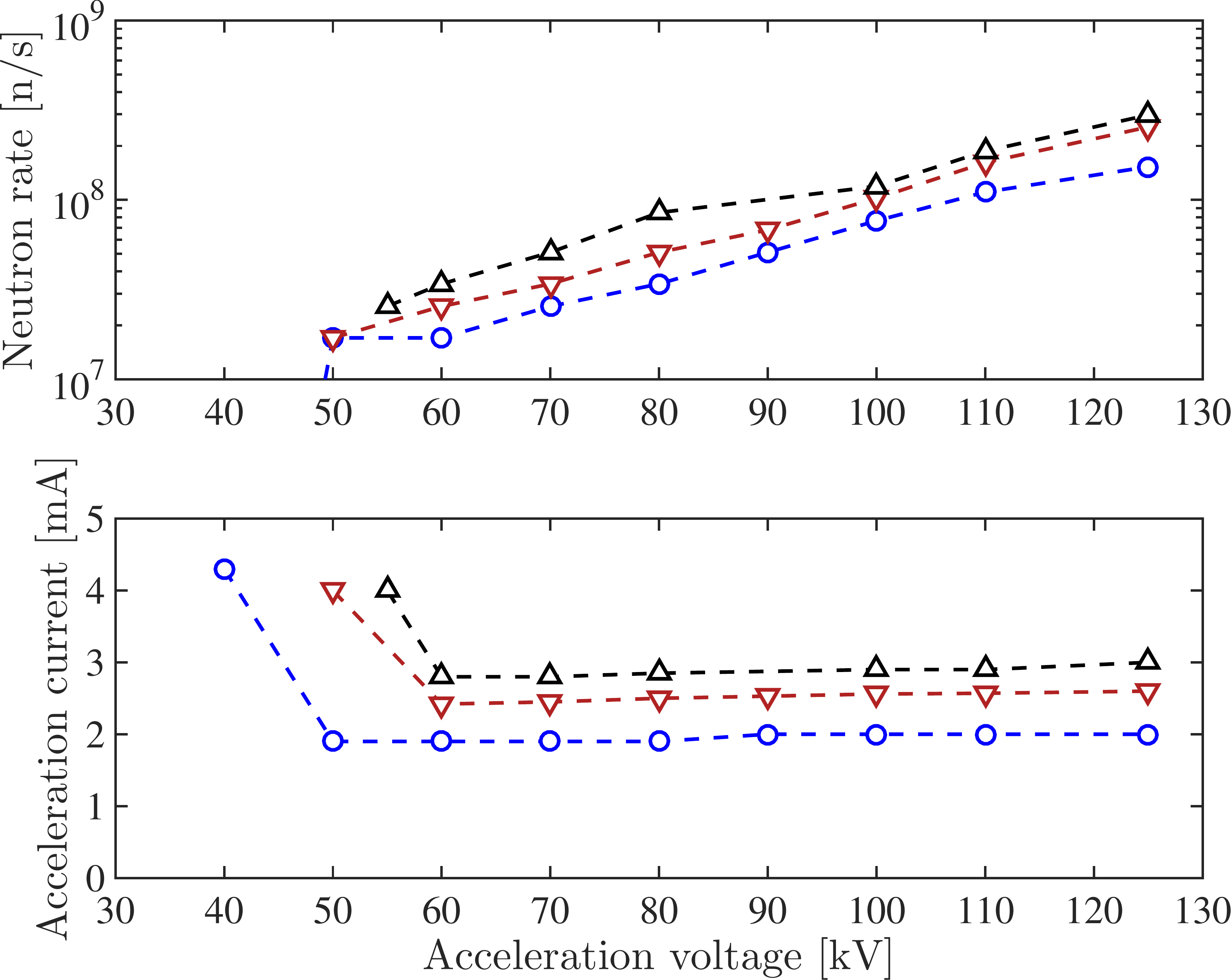}
        \vskip -0.1cm
        \caption{
            The measured neutron yield vs.\ acceleration voltage for the Brown DD108 neutron generator.
            The blue ($\bigcirc$), red ($\bigtriangledown$), and black ($\bigtriangleup$) curves represent data collected with the magnetron delivering 274~W, 376~W, and 496~W, respectively, to the deuterium plasma.
            The neutron rate shown in blue drops off scale at 40~kV in the top frame to 0~n/s. 
            The neutron generator was operating continuously (duty cycle of 1). 
            The plasma pressure was measured to be 4.3--4.4~mTorr for all three measurements.
            Data provided by Adelphi Technology, Inc.\ and produced here with permission~\cite{AdelphiTechnologyPrivate}.
            We estimate a factor of $\sim$2 uncertainty on the total neutron rate.
        }
        \vskip -0.5cm
        \label{fig:brown_dd108_params_yield_vs_accel_voltage}
    \end{center}
\end{figure}

The neutron output of the Brown DD108 was measured over a wide range of operating parameters by the vendor.
The measured neutron yield as a function of acceleration voltage is shown in Fig.~\ref{fig:brown_dd108_params_yield_vs_accel_voltage} for three different levels of power delivered to the deuterium plasma by the magnetron.
An order of magnitude of dynamic range in neutron yield can be obtained by adjusting the acceleration high voltage.
The increase in neutron flux is due to the increasing cross-section of the \dd{} reaction with deuterium ion energy.
The acceleration current, a measure of the number of deuterium ions on target, remains constant as the acceleration voltage is increased. 

\begin{figure}[!htb]
    \begin{center}
        \includegraphics[width=0.480\textwidth]{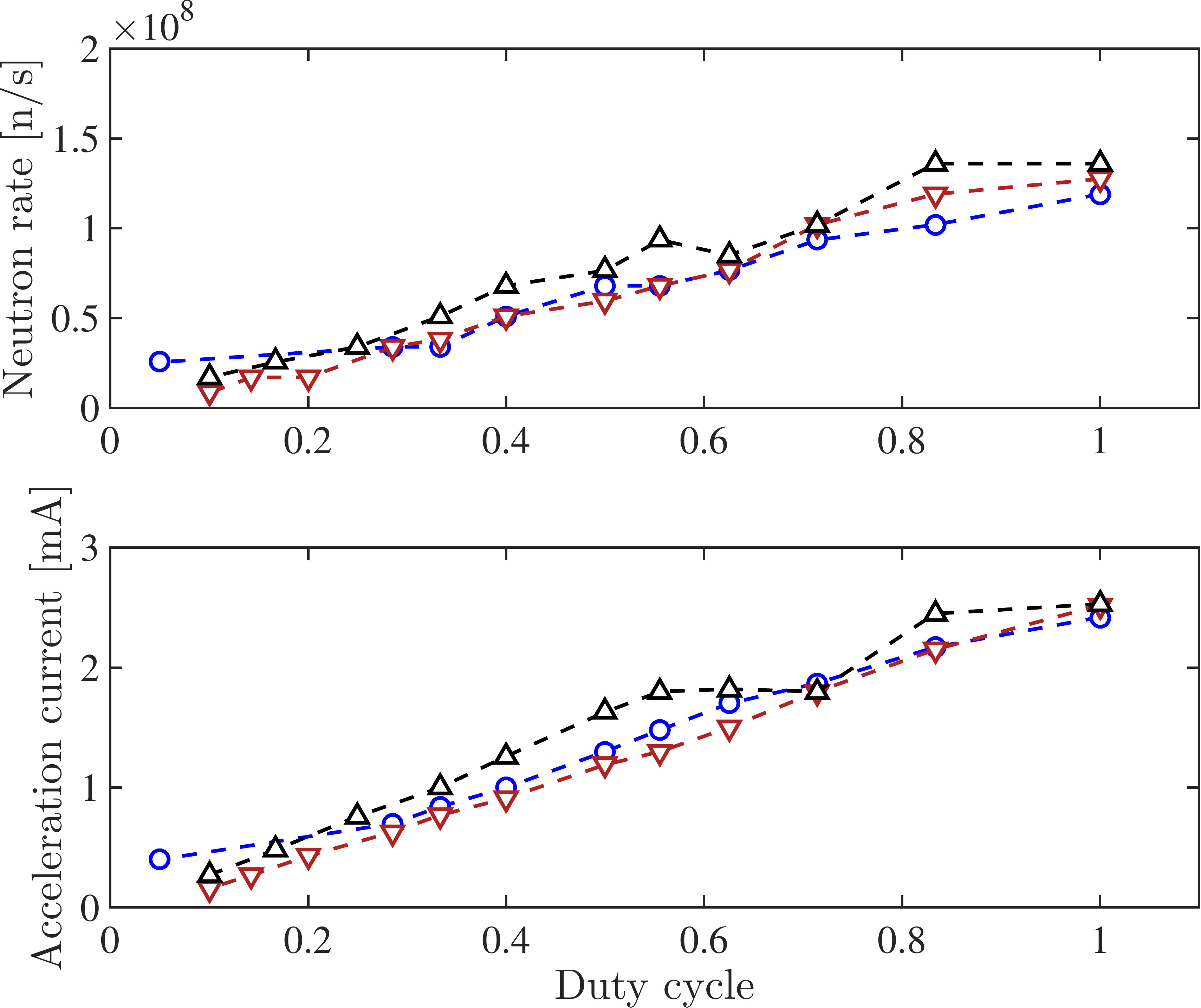}
        \vskip -0.1cm
        \caption{
            The measured neutron yield vs.\ duty cycle during pulsed operation for the Brown DD108 neutron generator.
            The blue ($\bigcirc$), red ($\bigtriangledown$), and black ($\bigtriangleup$) curves represent magnetron pulse widths of 5~ms, 1~ms, and 100~us, respectively.
            The other operating parameters were held approximately constant: $V_{\textrm{HV}} = 100$~kV, $I_{m} = 70$~mA.
            The power delivered to the deuterium plasma by the magnetron was measured to be 320--330~W during these measurements. 
            The plasma pressure for each of the three measurements was measured to be in the range 4.2--4.5~mTorr.
            Data provided by Adelphi Technology, Inc.\ and produced here with permission~\cite{AdelphiTechnologyPrivate}.
            We estimate a factor of $\sim$2 uncertainty on the total neutron rate.
        }
        \vskip -0.5cm
        \label{fig:brown_dd108_params_yield_vs_duty_cycle}
    \end{center}
\end{figure}

Pulsing of the neutron output is achieved by deuterium ion source control. 
Magnetron operation is pulsed for fine adjustment of neutron bunch width using a TTL control signal.
The pulse width and frequency can be arbitrarily varied to achieve the desired yield and time profile subject to the nominal minimum pulse width of 100~$\mu$s.
The measured neutron yield and acceleration current is shown in Fig.~\ref{fig:brown_dd108_params_yield_vs_duty_cycle} as a function of duty cycle for three distinct pulse width modes.
In contrast to Fig.~\ref{fig:brown_dd108_params_yield_vs_accel_voltage}, the acceleration current scales linearly with increasing duty cycle tracking the measured neutron flux.

\begin{figure}[!htb]
    \begin{center}
        \includegraphics[width=0.480\textwidth]{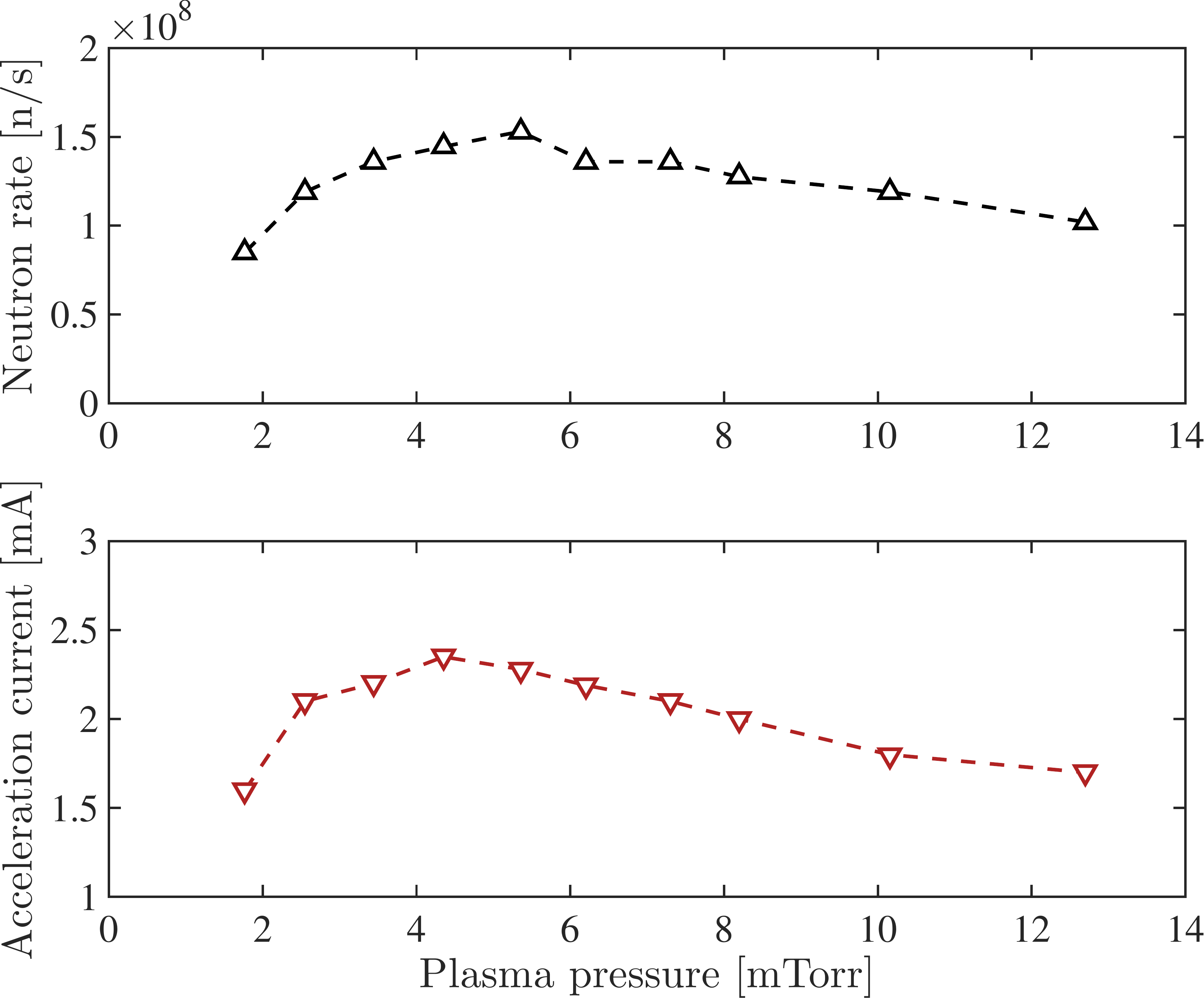}
        \vskip -0.1cm
        \caption{
            The measured neutron yield vs.\ plasma pressure for the Brown DD108 neutron generator.
            The black ($\bigtriangleup$) curve in the top frame shows the measured neutron flux as a function of plasma pressure.
            The red ($\bigtriangledown$) curve in the bottom frame shows the corresponding acceleration current.
            The neutron generator was operating continuously (duty cycle of 1). 
            The other operating parameters were held constant: $V_{\textrm{HV}} = 110$~kV, $I_{m} = 69$~mA.
            The power delivered to the deuterium plasma by the magnetron was 320~W.
            Data provided by Adelphi Technology, Inc.\ and produced here with permission~\cite{AdelphiTechnologyPrivate}.
            We estimate a factor of $\sim$2 uncertainty on the total neutron rate.
        }
        \vskip -0.5cm
        \label{fig:brown_dd108_params_yield_vs_plasma_pressure}
    \end{center}
\end{figure}

The neutron flux and acceleration current are shown as a function of deuterium plasma pressure in Fig.~\ref{fig:brown_dd108_params_yield_vs_plasma_pressure}. 
The maximum neutron yield is achieved at a plasma pressure of $\sim$5~mTorr.
There are fewer available deuterium ions at lower plasma pressures, which reduces the neutron flux.
At higher plasma pressures, the fraction of singly ionized molecular deuterium molecules (primarily D$_{2}^{+}$ and D$_{3}^{+}$) increases.
The energy provided by the acceleration potential is split between the atoms in molecular deuterium projectiles incident upon the target~\cite{IAEA2012}.
On average, the reduced energy per deuterium atom provides a lower cross-section for the nuclear \dd{} reaction and results in a lower neutron flux. 

\subsection{Time-of-flight experimental setup}

\begin{figure*}[!htb]
    \begin{center}
        \includegraphics[width=0.95\textwidth]{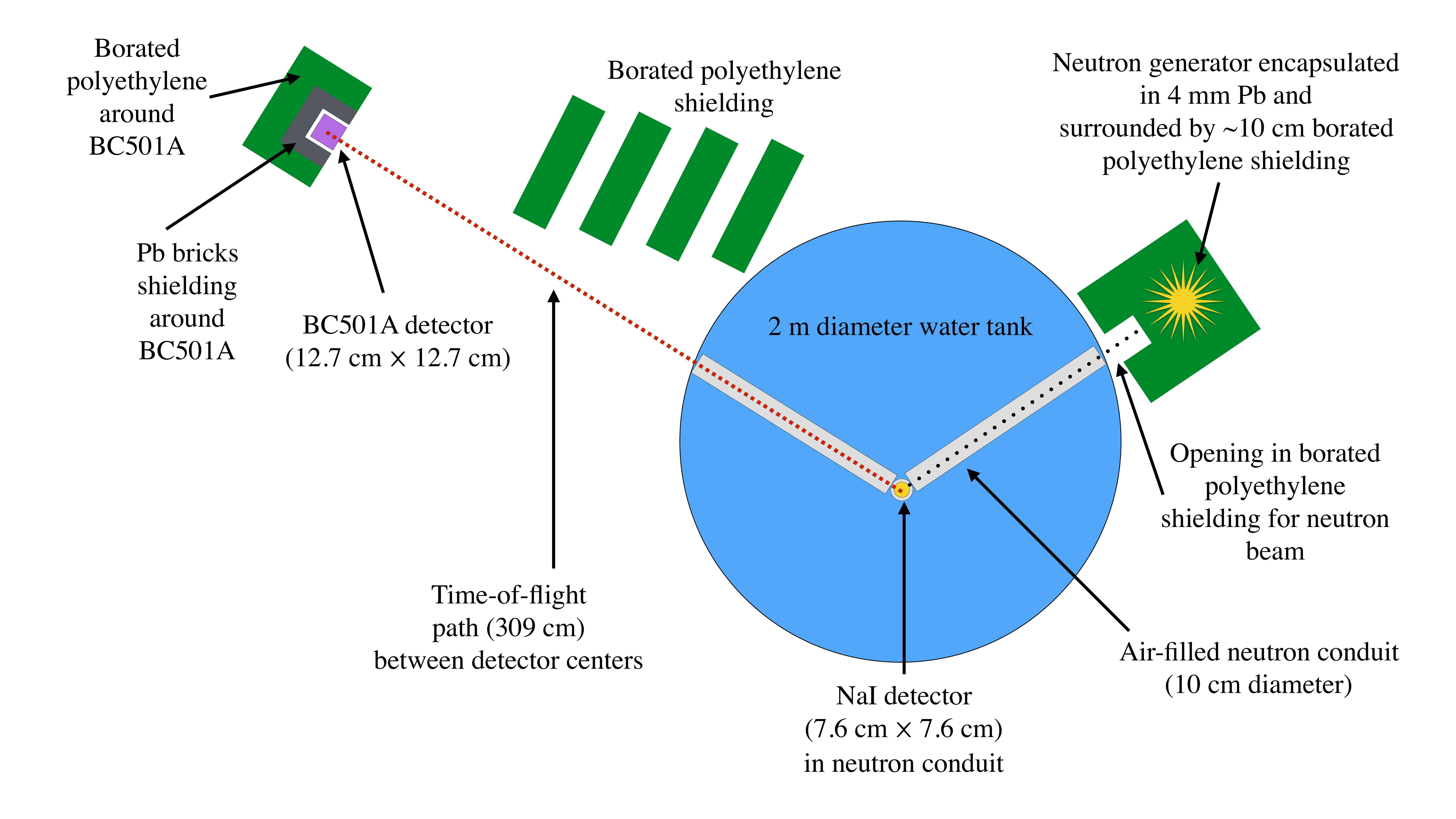}
        \vskip -0.1cm
        \caption{
            The experimental setup for the neutron ToF measurement performed at Brown University.
            The DD108 is shown at right encapsulated in borated polyethylene (green).
            The angled neutron collimation tube is depicted in gray inside the 2~m diameter water tank, with the 7.6~cm NaI(Tl) detector at the vertex (yellow circle).
            The far BC501A detector is shown in purple with surrounding Pb and borated polyethylene shielding.
            The incident neutrons from the generator accepted by the collimation path are represented by the black dotted line, and the 3~m ToF measurement path is shown by the red dashed line.
        }
        \vskip -0.5cm
        \label{fig:20150306_brown_neutron_generator_spectrum_experiment_diagram}
    \end{center}
\end{figure*}

The ToF experimental setup shown in Fig.~\ref{fig:20150306_brown_neutron_generator_spectrum_experiment_diagram} was used to assay the energy spectrum of neutrons produced by the DD108 neutron generator.
A similar experimental configuration has been used by others for studies of the NaI(Tl) nuclear recoil quenching factor~\cite{Chagani2008, Collar2013a}.

The neutron generator was encapsulated in $\sim$10~cm of borated polyethylene shielding with an opening to provide a beam of unmoderated neutrons.
A 4~mm-thick Pb sheet was used to suppress bremsstrahlung x-rays produced by the device.
A 10~cm diameter air-filled conduit was submerged in a 2~m diameter water tank to provide a kinked collimation path subtending an angle of 114$^{\circ}$.
This angled air-filled conduit enforced a scattering angle of $66^{\circ} \pm 4^{\circ}$ for neutrons following the collimation path through the water tank.
A 7.6~cm diameter, 7.6~cm tall NaI(Tl) detector (Ludlum 44-20) was installed inside the vertex of the air-filled conduit to provide a $t_{0}$ for the ToF measurement.
The water tank also functioned to reduce accidental coincidence backgrounds by shielding the NaI(Tl) detector from ambient gamma rays.
A Bicron 501A (BC501A) liquid scintillator (12.7~cm diameter, 12.7~cm height) detector was placed outside the water tank in line with the second leg of the collimation path.

The average ToF path was measured to be $309 \pm 4$~cm from the center of the NaI(Tl) detector to the center of the BC501A. 
Coincident events in the NaI(Tl) and BC501A detectors were used to characterize the energy spectrum of neutrons produced by the DD108 by measuring the particle ToF between the two detectors.
The BC501A was positioned to ensure $>$1~m of water shielding between the DD108 and BC501A to suppress accidental coincidences due to line-of-sight neutrons from the generator interacting in the far detector.
The face of the BC501A detector in line with the beam path was left unshielded to increase the efficiency of detection of neutrons from the true ToF path. 
All other sides of the BC501A detector were shielded by $\sim$5 cm of Pb to reduce the accidental coincidence rate from ambient gamma rays interacting in the BC501A.
A $\sim$5 cm layer of borated polyethylene was constructed outside of the BC501A Pb shield to reduce the false coincidence rate produced by unwanted neutron shine off passive surfaces in the room.

The ideal signal events consist of a neutron leaving the neutron generator, scattering once in the NaI(Tl) detector, and then scattering in the far BC501A detector without scattering in passive materials during the journey.
The deposited energy $E_{\textrm{nr},\textrm{Na}}$ from neutron scatters off Na in the NaI(Tl) detector is given by Eq.~\ref{eq:recoil_energy_equation}, where $m_{A}$ is the atomic mass of Na.
The neutron velocity is obtained by measuring the ToF between the NaI(Tl) and BC501A detectors.
The energy of each neutron can be directly determined from its velocity as $E_{n,\textrm{meas}} = 1/2mv^{2}$.
Neutrons with the nominal expected mean energy for our experimental setup of 2.45~MeV are non-relativistic, traveling at ~7\% the speed of light.
It takes these neutrons 46~ns to travel 1~m.
The measured neutron energy using ToF between the two detectors, $E_{n,\textrm{meas}}$, is lower than the energy of the neutron incident on the NaI(Tl) detector, $E_{n}$, due to the energy deposited in the NaI(Tl).
Eq.~\ref{eq:tof_energy} and Eq.~\ref{eq:tof_inc_energy} are used to account for the lost recoil energy assuming Na recoils, $E_{\textrm{nr},\textrm{Na}}$, and reconstruct $E_{n}$ given $E_{n,\textrm{meas}}$:

\begin{equation} \label{eq:tof_energy}
    E_{n} = E_{n,\textrm{meas}} + E_{\textrm{nr},\textrm{Na}} \, \text{.}
\end{equation}

The true measured incident energy is given by

\begin{equation} \label{eq:tof_inc_energy}
    E_{n} = \frac{E_{n,\textrm{meas}}}{1-\zeta} \, \text{,}
\end{equation}

\noindent
where $m_{A}$ is the atomic mass of Na and $\zeta$ is given in Eq.~\ref{eq:tof_zeta}.
Events due to neutrons that scatter multiple times in the NaI(Tl) crystal contribute to a featureless ToF background that does not affect the determination of the single-scatter peak parameters~\cite{Chagani2008, Collar2013a}.
The experimental setup was not sensitive to elastic iodine recoils in the NaI(Tl) detector due to the lower energy transfer to these nuclei as expected from Eq.~\ref{eq:recoil_energy_equation} and the lower nuclear recoil signal yield for iodine.
There are several inelastic recoil modes for iodine, only one of which remains after the analysis cuts described in Sec.~\ref{sec:n_tof_measurement_analysis}.
The remaining mode is $^{127}$I(n,\,n$^{\prime} \gamma$) producing a 57.6~keV gamma ray also seen in Ref.~\cite{Collar2013a}.

After $\times$10 amplification, the NaI(Tl) and BC501A signals were digitized at 1~GHz in Ch1 and Ch2 of an 8~bit Lecroy LT583 oscilloscope in sequence mode.
The scope was externally triggered based upon the overlap coincidence of a 400 ns gate pulse from the NaI(Tl) and a 200 ns gate pulse from the BC501A.
A discriminator was used to set hardware thresholds of $\sim$20~mV and $\sim$150~mV for the NaI(Tl) and BC501A signals, respectively, for the signal heights as measured at the digitizer.
Each sequence of 50~triggers was pulled from the oscilloscope to a control computer via Ethernet and saved to disk.
This coincidence setup provides a trigger regardless of signal arrival order from the two detectors, which allows verification of the expected flat accidental coincidence background.

\subsection{Measurement of the neutron time-of-flight spectrum} \label{sec:n_tof_measurement_analysis}

We provide a detailed overview of the analysis process and report results for Target Orientation A in Sec.~\ref{sec:tof_orientation_a}.
The same analysis process was repeated for Target Orientation B, and the results are summarized in Sec.~\ref{sec:tof_orientation_b}.
Identical cuts and algorithms were used for the analysis of datasets for both DD108 target orientations.
More detail is available in Ref.~\cite{Verbus2016}.

\subsubsection{DD108 target orientation A} \label{sec:tof_orientation_a}

A total of $2.5\times10^{5}$ coincidence triggers were acquired in this configuration and used for the analysis.
The $t_{0}$ of every NaI(Tl) and BC501A pulse was determined by the point where the pulse rose to 10\% of its maximum value.
The time difference between the $t_{0}$ of the pulses in each coincident event was used to measure the ToF.
This is referred to as ``raw ToF'' in the text.
The offset due to pulse shape differences in NaI(Tl) and BC501A was calibrated out using the raw ToF location of the gamma ray coincidence peak.
The calibrated time scale is referred to as ``corrected ToF.'' 
Basic data quality cuts were applied.
The data quality cuts have a total combined acceptance of 84.5\% of all acquired events.

\begin{figure}[!htb]
    \begin{center}
        \includegraphics[width=0.480\textwidth]{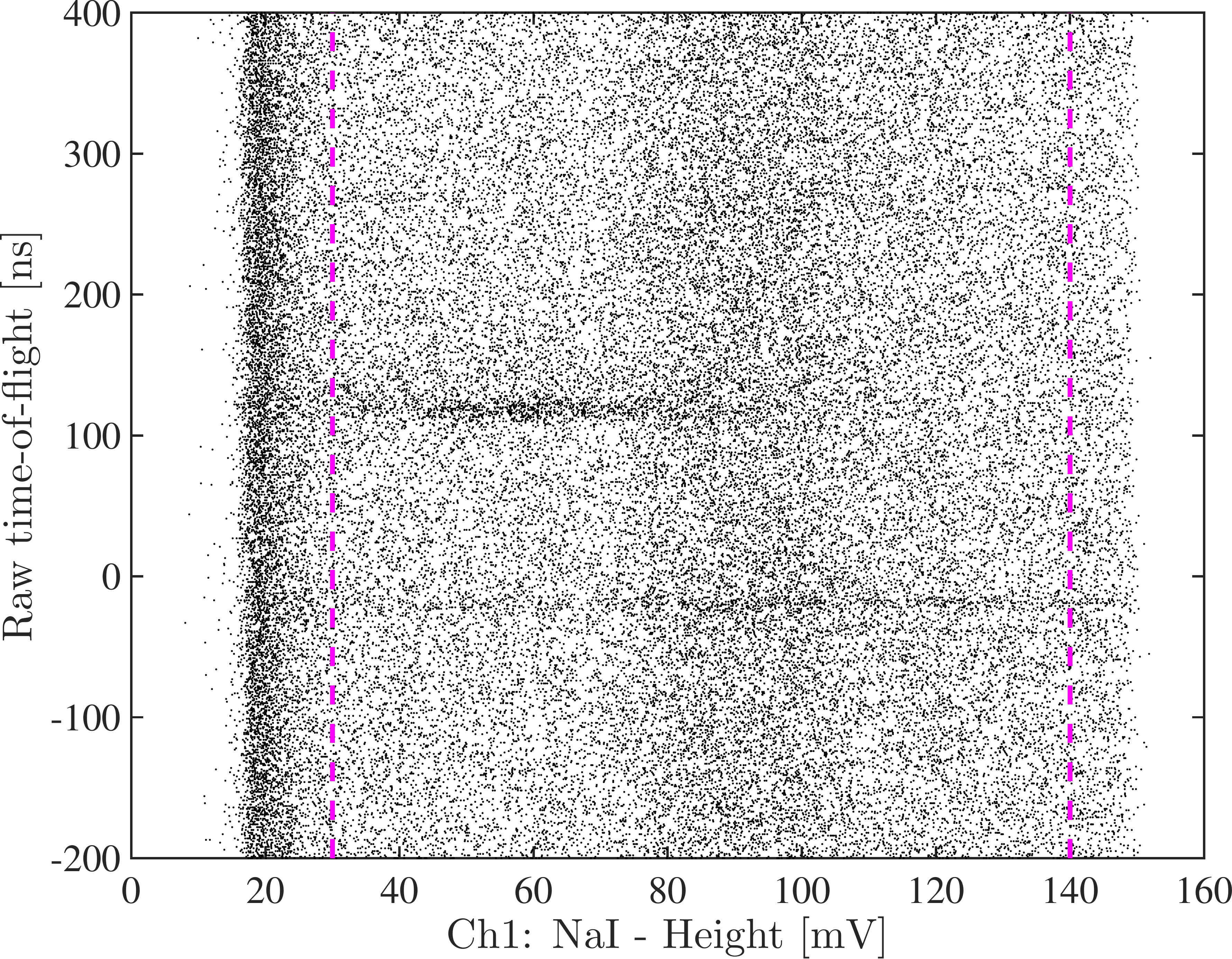}
        \vskip -0.1cm
        \caption{
            Target Orientation A.
            The raw ToF vs.\ NaI(Tl) pulse height distribution is shown for events passing the area and data quality cuts.
            The lower and upper analysis thresholds at 30 and 140~mV, respectively, are represented by the vertical dashed magenta lines.
            This figure is produced before correcting the ToF based upon the known gamma ray propagation time between detectors.
        }
        \vskip -0.5cm
        \label{fig:dd_neutron_tof_orientation_a_tof_vs_nai_height_scatter}
    \end{center}
\end{figure}

A pulse height cut was applied to ensure pulses from the NaI(Tl) detector were between 30--140~mV, as measured at the oscilloscope.
The limits on NaI(Tl) pulse size reduce contamination from background gamma rays while maintaining high efficiency for neutron scatters producing coincident events in both detectors as can be seen in Fig.~\ref{fig:dd_neutron_tof_orientation_a_tof_vs_nai_height_scatter}.
Neutrons produced by the \dd{} source that scatter in both the NaI(Tl) and BC501A detectors are visible in the horizontal band at $\sim$115~ns.
Residual background gamma ray events that produce signals in both the NaI(Tl) and BC501A detectors are represented in the horizontal band at roughly -20~ns.
The vertical band of accidental coincidences at $\sim$20~mV is just above the discriminator threshold.
The vertical band observed between 80 and 100~mV is produced by 57.6~keV gamma rays from $^{127}$I(n,\,n$^{\prime} \gamma$) inelastic scatters.

\begin{figure}[!htb]
    \begin{center}
        \includegraphics[width=0.480\textwidth]{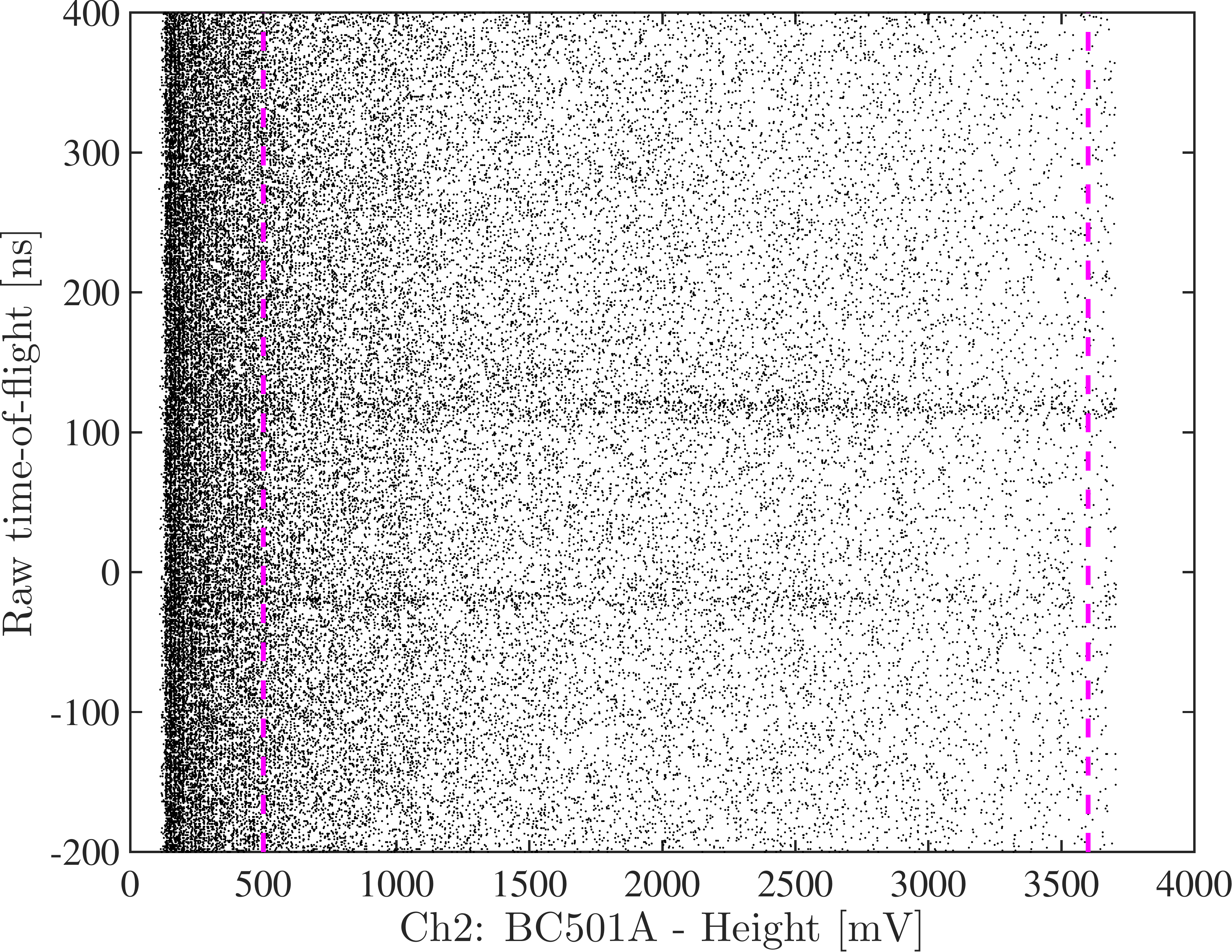}
        \vskip -0.1cm
        \caption{
            Target Orientation A.
            The raw ToF vs.\ BC501A pulse height distribution for events passing the area and data quality cuts.
            The lower and upper analysis thresholds at 500 and 3600~mV, respectively, are represented by the vertical dashed magenta lines.
            This figure is produced before correcting the ToF based upon the known gamma ray propagation time between detectors.
        }
        \vskip -0.5cm
        \label{fig:dd_neutron_tof_orientation_a_tof_vs_bc501_height_scatter_ch2}
    \end{center}
\end{figure}

A pulse height cut was applied to ensure pulses from the BC501A detector were between 500 and 3600 mV, as measured at the scope.
The cut bounds the BC501A pulse height in Ch2 were set to ensure effective discrimination on the low end while avoiding the saturation on the high end.
The raw ToF vs.\ BC501A pulse pulse height in Ch2 is shown in Fig.~\ref{fig:dd_neutron_tof_orientation_a_tof_vs_bc501_height_scatter_ch2}.

\begin{figure}[!htb]
    \begin{center}
        \includegraphics[width=0.480\textwidth]{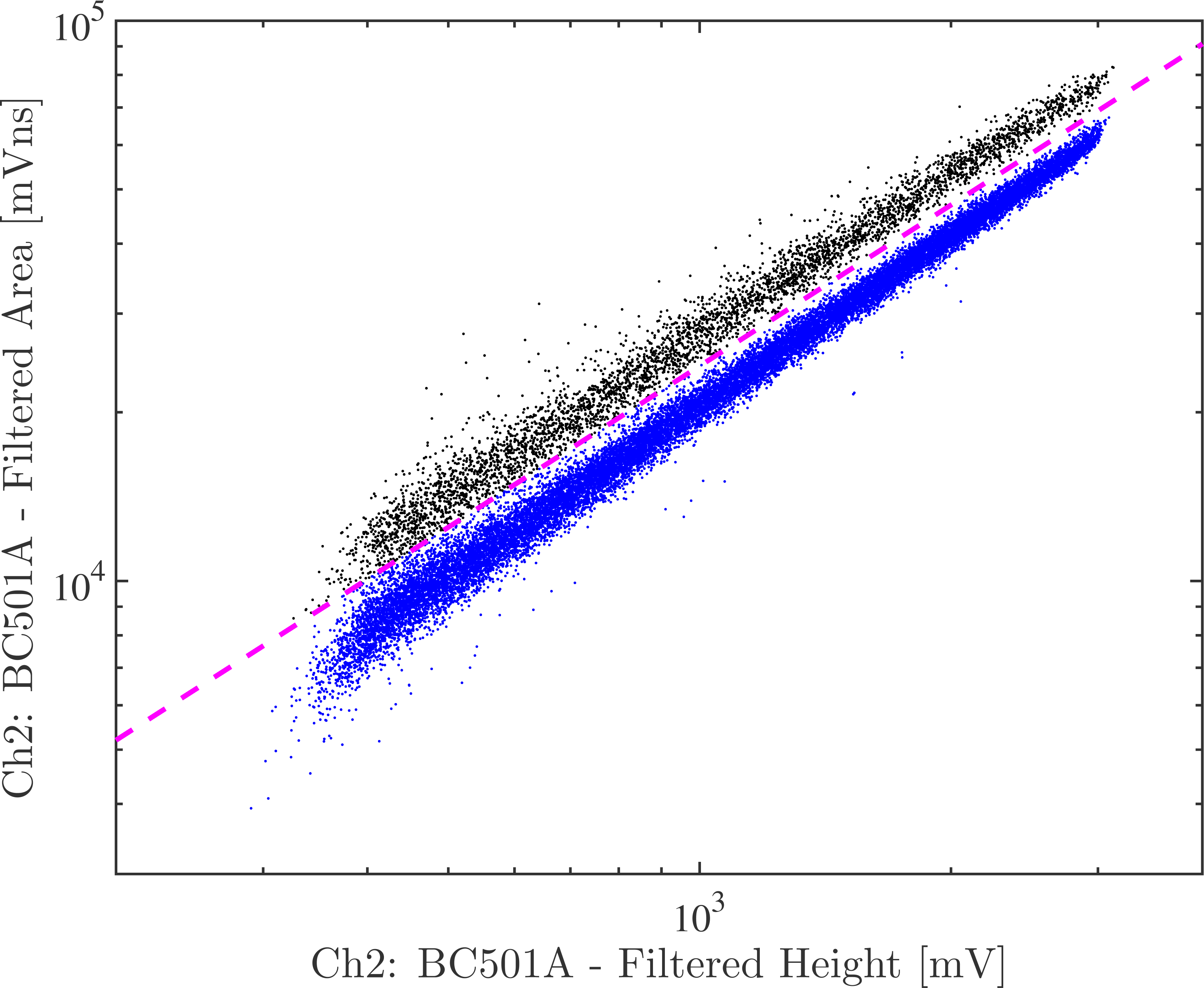}
        \vskip -0.1cm
        \caption{
            Target Orientation A.
            The BC501A discrimination decision boundary in the area vs.\ height parameter space for events passing all cuts is represented by the dashed magenta line.
            Gamma ray events are depicted in blue while neutron events are depicted in black.
            The decision boundary is given by $y = 33 x^{0.955}$.
        }
        \vskip -0.5cm
        \label{fig:dd_neutron_tof_orientation_a_bc501_area_vs_height_discrimination}
    \end{center}
\end{figure}

The pulse-shape discrimination (PSD) capabilities of the BC501A detector were used to differentiate between neutron and gamma ray events passing all other cuts, with the results shown in Fig.~\ref{fig:dd_neutron_tof_orientation_a_bc501_area_vs_height_discrimination}.
The event traces were smoothed using a low-pass Butterworth filter with a cutoff frequency of 50 MHz before determining the area and height of each pulse.
The resulting quantities are referred to as filtered area and filtered height.
The filtered quantities are used for PSD only.

\begin{figure}[!htb]
    \begin{center}
        \includegraphics[width=0.480\textwidth]{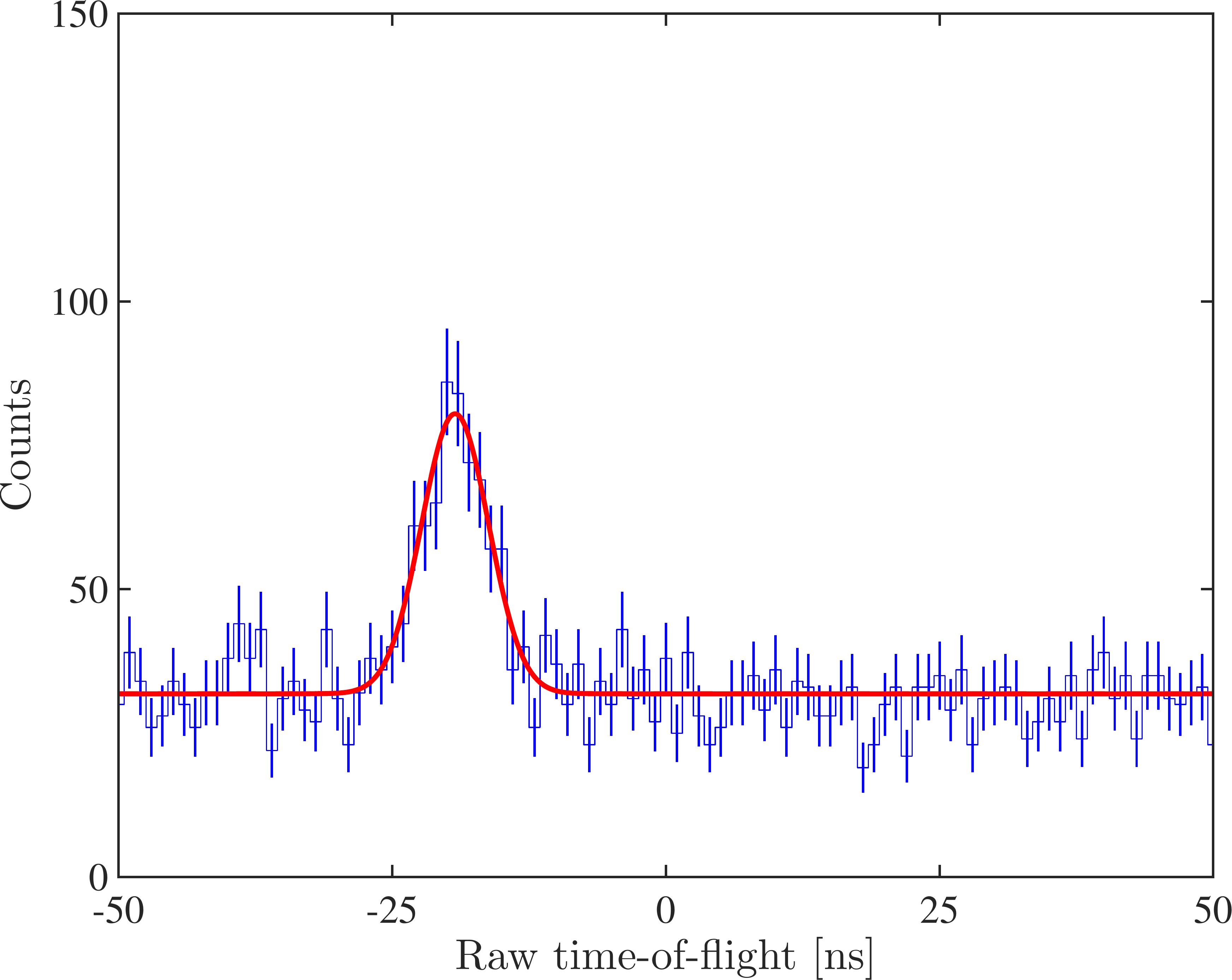}
        \vskip -0.1cm
        \caption{
            Target Orientation A. 
            The Gaussian fit to the gamma ray ToF spectrum is indicated by the solid red line.
            The gamma ray ToF was measured to be -19.3 $\pm$ 0.3~ns with a measured sigma of 3.1 $\pm$ 0.3~ns.
            Uncertainties are statistical.
            The fit region has $\chi^{2}$/dof = 83.9/95 yielding a p-value of 0.78.
            Bins at the extremes of the fit domain with an expectation of fewer than 10 counts were combined when calculating $\chi^{2}$.
        }
        \vskip -0.5cm
        \label{fig:dd_neutron_tof_orientation_a_histogram_gamma_peak_fit_zoom}
    \end{center}
\end{figure}

\begin{figure}[!htb]
    \begin{center}
        \includegraphics[width=0.480\textwidth]{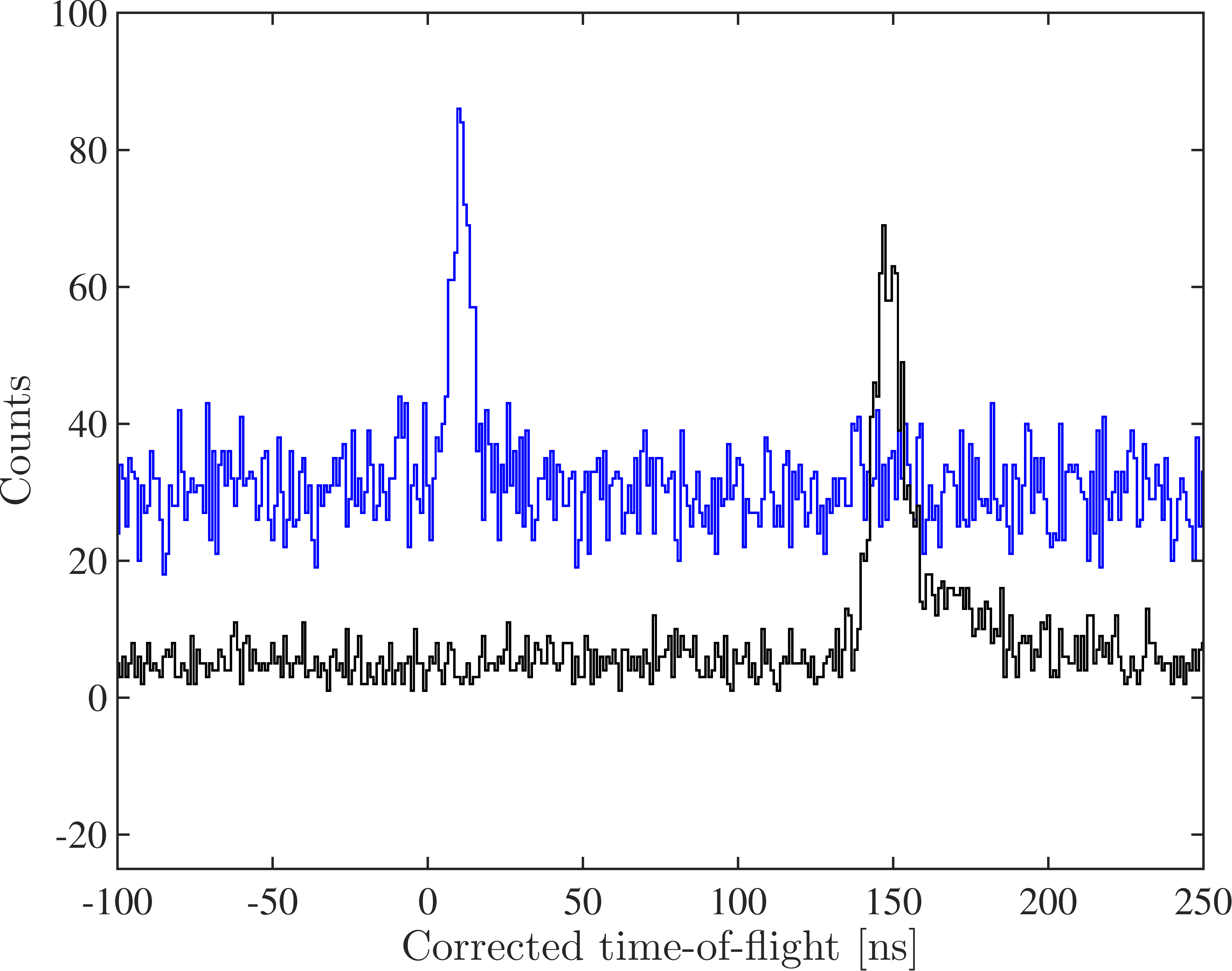}
        \vskip -0.1cm
        \caption{
            Target Orientation A.
            The individual ToF spectra for gamma ray (blue) and neutron (black) events passing all cuts are shown as selected in Fig.~\ref{fig:dd_neutron_tof_orientation_a_histogram_gamma_peak_fit_zoom}.
            The ToF axis has been calibrated using the Gaussian fit to the gamma ray peak and the known gamma ray propagation time of 10.3~ns between detectors.
        }
        \vskip -0.5cm
        \label{fig:dd_neutron_tof_orientation_a_histogram_of_individual_tof_spectra}
    \end{center}
\end{figure}

A Gaussian was fit to the gamma ray coincidence peak obtained after selecting electron recoil events in the BC501A to obtain the $t_{0}$ calibration, as shown in Fig.~\ref{fig:dd_neutron_tof_orientation_a_histogram_gamma_peak_fit_zoom}.
The measured raw ToF values are corrected using this calibration of the location of the gamma ray peak and the expected 10.3~ns gamma ray ToF between the NaI and BC501A.
The calibration of the ToF scale using gamma ray coincidences corrects for any unwanted time offset between the NaI(Tl) and BC501A channels due to cable lengths, signal delays in electronics, and, most significantly, the variation in the algorithmic determination of pulse start time $t_{0}$ for the signals provided by the NaI(Tl) and BC501A detectors.
To reduce the systematic uncertainty due to the different pulse shapes for neutron and gamma ray interactions the algorithm used to determine $t_{0}$ for each pulse is based upon the identification of a constant fraction of the pulse height.
This is algorithm is primarily sensitive to the rise time of the pulse, while particle type primarily influences the long tail after the prompt scintillation component~\cite{Knoll2000}. 
The measured variance of the gamma ray coincidence peak provides an estimate of the contribution to the intrinsic ToF resolution from detector size, angular acceptance, electronics, and analysis algorithms.
The corrected ToF distributions for neutron and gamma ray events are shown in Fig.~\ref{fig:dd_neutron_tof_orientation_a_histogram_of_individual_tof_spectra}.

A non-Gaussian tail at high ToF due neutron energy loss in passive material has been noted in other similar neutron scattering experiments~\cite{Chagani2008, Collar2013a}.
To accommodate the expected high ToF tail, the modified Crystal Ball function in Eq.~\ref{eq:crystal_ball_function} was fit to the observed neutron corrected ToF spectrum~\cite{Oreglia1980, Gaiser1982, Skwarnicki1986, Santos2014}. 
The Crystal Ball function is a smooth function composed of a Gaussian stitched together with a power law tail:

\begin{equation} \label{eq:crystal_ball_function}
    y = 
    \begin{cases} 
        N \exp{\left[ \frac{-(x - \mu)^{2}}{2 \sigma^{2}} \right]} + C \, \text{,} & \text{if } \frac{x - \mu}{\sigma} < -\alpha \, \text{,} \hfill \\ \\
        N \frac{ \left( \frac{n}{\lvert \alpha \rvert} \right)^{n} \exp{\left( \frac{-\alpha^{2}}{2} \right)} }{ \left( \frac{n}{\lvert \alpha \rvert} - \lvert \alpha \rvert + \frac{x - \mu}{\sigma} \right)^{n} } + C \, \text{,} & \text{if } \frac{x - \mu}{\sigma} \geq -\alpha \, \text{.} \hfill 
    \end{cases} 
\end{equation}

\noindent
We modified the signs and inequalities to produce a tail at high ToF, rather than low ToF.
The Gaussian mean and width are given by $\mu$ and $\sigma$, respectively.
The parameter $\alpha$ controls location of transition from the Gaussian to the power law tail.
The parameter $n$ controls the slope of the power law, and $N$ is an arbitrary overall scaling factor.
We accommodate the flat accidental coincidence background with the parameter $C$.

\begin{figure}[!htb]
    \begin{center}
        \includegraphics[width=0.480\textwidth]{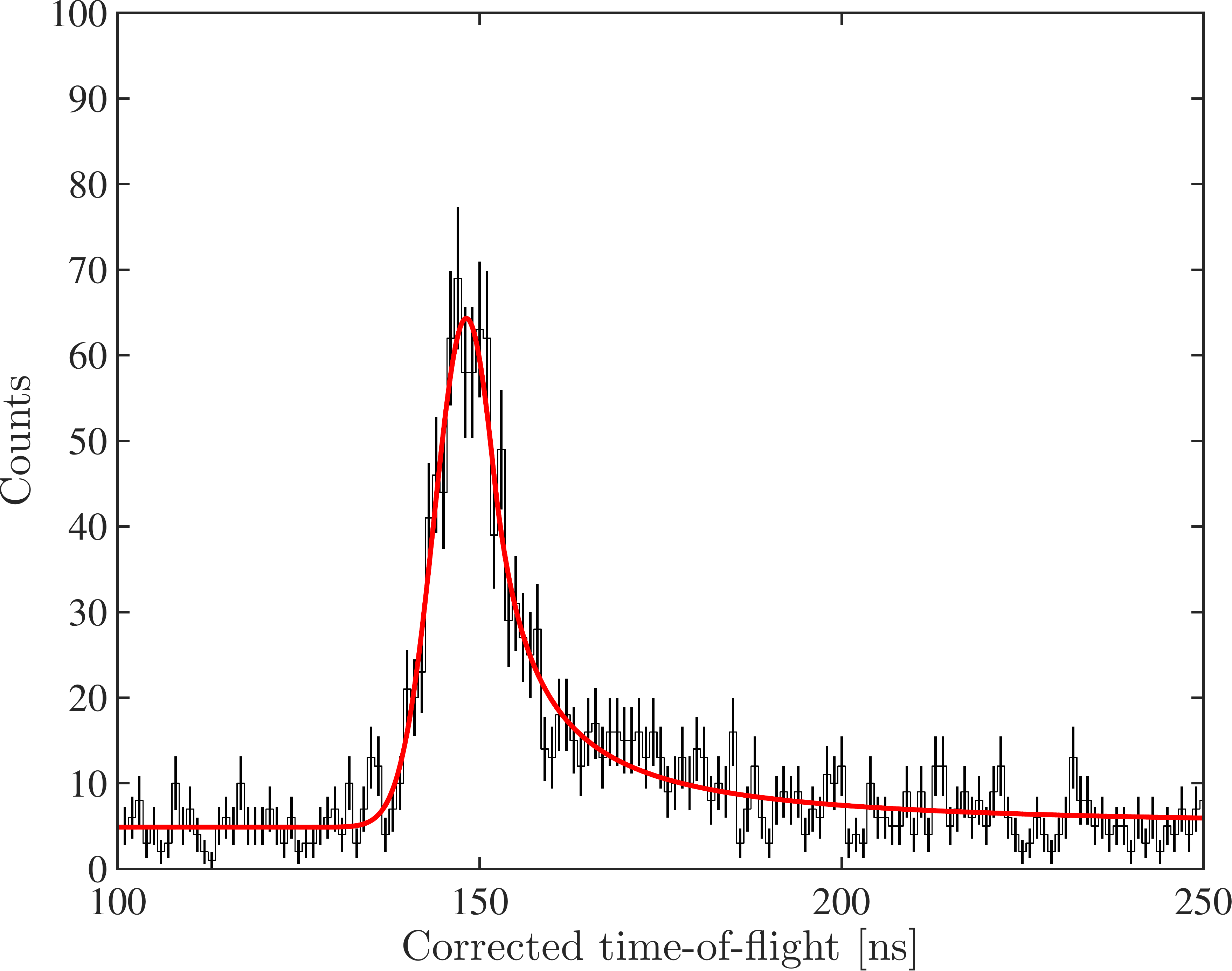}
        \vskip -0.1cm
        \caption{
            Target Orientation A.
            The Crystal Ball function fit to the neutron ToF spectrum is indicated by the solid red line.
            The fit region has $\chi^{2}$/dof = 23.7/25 yielding a p-value of 0.54.
            Bins at the extremes of the fit domain with an expectation of fewer than 10 counts were combined when calculating $\chi^{2}$.
        }
        \vskip -0.5cm
        \label{fig:dd_neutron_tof_orientation_a_histogram_neutron_peak_fit_zoom}
    \end{center}
\end{figure}

This functional form provides a reproducible, algorithmic determination of the location of the transition between underlying Gaussian neutron energy spectrum produced by the DD108 and the lossy tail of events at higher corrected ToF.
The Gaussian mean and variance parameters in the Crystal Ball function fit shown in Fig.~\ref{fig:dd_neutron_tof_orientation_a_histogram_neutron_peak_fit_zoom} were used to characterize the underlying neutron energy spectrum from the \dd{} neutron generator.

The mean neutron corrected ToF was measured to be $148.2 \pm 0.4$~ns with a resolution ($\sigma/\mu$) of 3\%.
Eq.~\ref{eq:tof_inc_energy} combined with $E_{n,\textrm{meas}} = 1/2 m v^{2}$ provides a mean neutron energy produced by the \dd{} source of 2.401~$\pm$~0.012~(stat)~$\pm$~0.060~(sys)~MeV.
The total systematic uncertainty has contributions from the uncertainties in the propagation distance between the detectors, the fixed angle of scatter, the angular acceptance of the collimation tubes, and most significantly the finite detector size and position.
The systematic uncertainty due to the choice of several analysis parameters was estimated by varying these parameters and repeating the analysis.
The systematic uncertainty due to the choice histogram bin width was estimated by repeating the analysis using 2~ns wide bins instead of the default 1~ns wide bins.
The systematic uncertainty due to the choice of fit region was estimated by expanding the neutron ToF fit region from 100--250~ns to 50--300~ns.
The systematic uncertainty due to position of the 140~mV NaI(Tl) pulse height cut was estimated by repeating the analysis using an upper pulse height cut of 80~mV/ns.
This alternative upper NaI(Tl) pulse height cut was chosen to remove the majority of $^{127}$I(n,\,n$^{\prime} \gamma$) events.
All uncertainties are reported in Table~\ref{tab:target_orientation_a_uncertainties_mean}.

\medskip
\begin{table}[htbp]
    \centering
    \caption{
        Target Orientation A.
        The statistical and systematic uncertainties on the mean energy of neutrons produced by the DD108 neutron generator are shown in columns two and three, respectively.
    }
    \label{tab:target_orientation_a_uncertainties_mean}
    \begin{tabular*}{\columnwidth}{@{\extracolsep{\fill}} lSS}
        \toprule
        Source of Uncertainty & {Statistical} & {Systematic} \\
        & {[\%]} & {[\%]} \\
        \midrule
        n and $\gamma$ peak fits & 0.5 & {-} \\
        Detector position & {-} & 2.4 \\
        Scattering angle & {-} & 0.5 \\
        Choice of bin width & {-} & 0.6 \\
        Choice of fit region & {-} & 0.02 \\
        NaI(Tl) upper area cut & {-} & 0.04 \\
        \midrule
        Total & 0.5 & 2.5 \\
        \bottomrule
    \end{tabular*}
\end{table}
\medskip

\medskip
\begin{table}[htbp]
    \centering
    \caption{
        Target Orientation A.
        The statistical and systematic uncertainties on the standard deviation of the neutron energy distribution produced by the DD108 neutron generator are shown in columns two and three, respectively.
    }
    \label{tab:target_orientation_a_uncertainties_sigma}
    \begin{tabular*}{\columnwidth}{@{\extracolsep{\fill}} lSS}
        \toprule
        Source of Uncertainty & {Statistical} & {Systematic} \\
        & {[\%]} & {[\%]} \\
        \midrule
        n and $\gamma$ peak fits & 13 & {-} \\
        Detector position & {-} & 2.4 \\
        Scattering angle & {-} & 0.5 \\
        Choice of bin width & {-} & 7 \\
        Choice of fit region & {-} & 1.0 \\
        NaI(Tl) upper area cut & {-} & 18 \\
        \midrule
        Total & 13 & 19 \\
        \bottomrule
    \end{tabular*}
\end{table}
\medskip

The measured variance of the gamma ray ToF distribution shown in Fig.~\ref{fig:dd_neutron_tof_orientation_a_histogram_gamma_peak_fit_zoom} provides an estimate of the contribution from our our experimental setup to the observed resolution. 
This experimental contribution was subtracted from the neutron ToF distribution variance to provide the most accurate determination of the intrinsic variance of the neutron energy distribution produced by the DD108. 
The fit determination of $\alpha$, the transition between Gaussian and power law tail, in the Crystal Ball function is correlated with the parameter estimate of $\sigma$, the standard deviation of the Gaussian component. 
The additional uncertainty due to this correlation is included in the reported statistical uncertainty for $\sigma$.
The standard deviation of the energy distribution of neutrons produced by the DD108 was measured to be 0.105~$\pm$~0.014~(stat)~$\pm$~0.020~(sys)~MeV after subtraction of the gamma ray peak variance.
The uncertainties are reported in Table~\ref{tab:target_orientation_a_uncertainties_sigma}.
The corresponding $\sigma/\mu$ of the neutrons produced by the \dd{} generator is 4.4\%~$\pm$~0.6\%~(stat)~$\pm$~0.8\%~(sys).

\subsubsection{DD108 target orientation B} \label{sec:tof_orientation_b}

A total of $5\times10^{5}$ coincidence triggers were acquired in this configuration and used for the analysis.
The cuts and analysis steps are identical to those in Sec.~\ref{sec:tof_orientation_a}.
The same data quality cuts were applied to the data as used in Sec.~\ref{sec:tof_orientation_a}.
The data quality cuts have a total combined acceptance of 89.7\% of all acquired events.
A more detailed account of the Target Orientation B results (similar to those presented here for Target Orientation A) is available in Ref.~\cite{Verbus2016}. 

\begin{figure}[!htb]
    \begin{center}
        \includegraphics[width=0.480\textwidth]{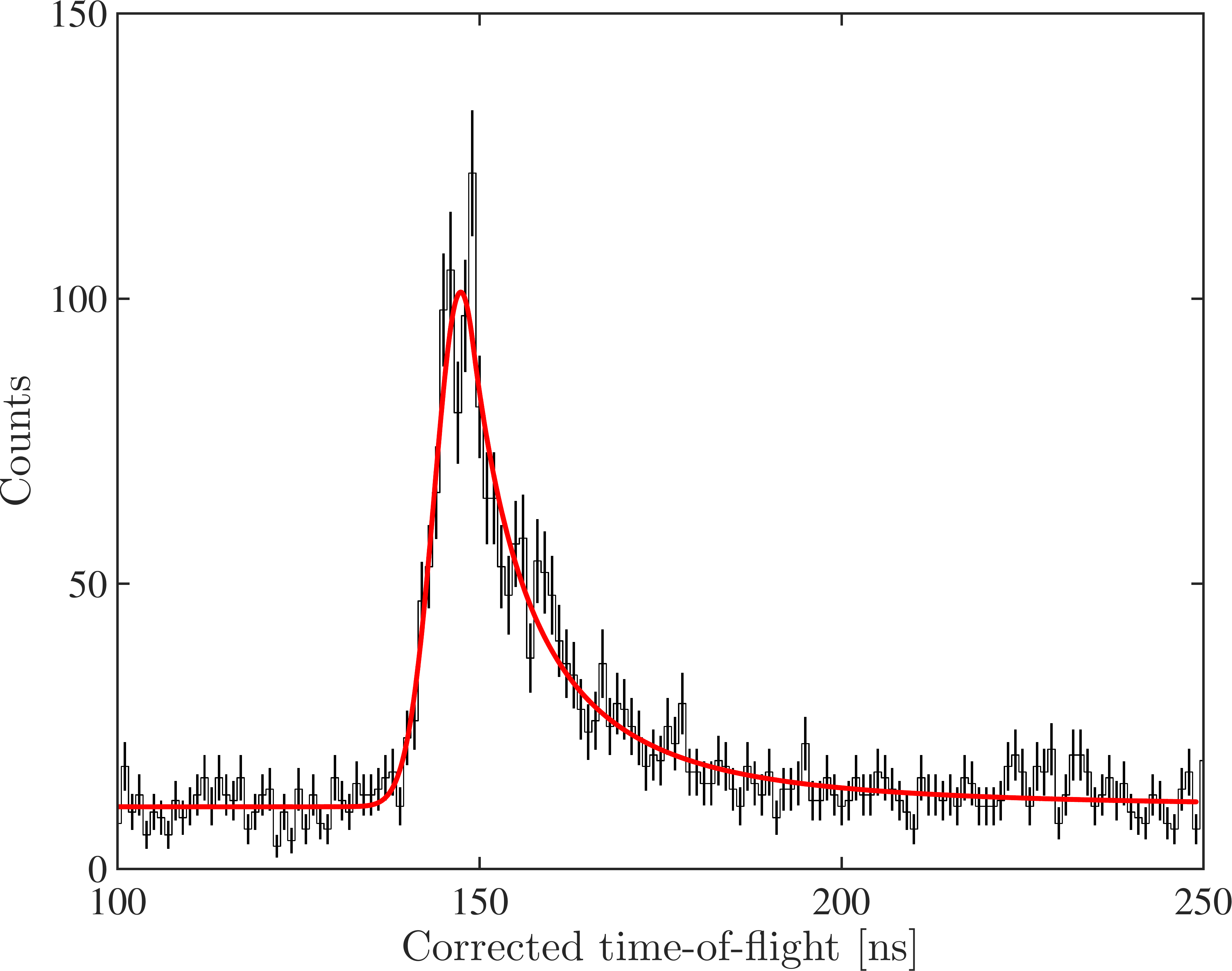}
        \vskip -0.1cm
        \caption{
            Target Orientation B.
            The Crystal Ball function fit to the neutron ToF spectrum is indicated by the solid red line.
            The fit region has $\chi^{2}$/dof = 159.3/143 yielding a p-value of 0.17.
            Bins at the extremes of the fit domain with an expectation of fewer than 10 counts were combined when calculating $\chi^{2}$.
        }
        \vskip -0.5cm
        \label{fig:dd_neutron_tof_orientation_b_histogram_neutron_peak_fit_zoom}
    \end{center}
\end{figure}

The final corrected ToF spectrum for neutron events passing all cuts is shown in Fig.~\ref{fig:dd_neutron_tof_orientation_b_histogram_neutron_peak_fit_zoom}.
The mean neutron corrected ToF was measured to be $147.4 \pm 0.4$~ns with a resolution of 2.5\%.
This corresponds to a measured neutron energy of 2.426~$\pm$~0.013~(stat)~$\pm$~0.08~(sys)~MeV incident on the NaI(Tl) detector.
The systematic uncertainties are calculated identically to Sec.~\ref{sec:tof_orientation_a} and are shown in Table~\ref{tab:target_orientation_b_uncertainties_mean}.
The measured mean of the neutron energy spectrum produced using Target orientation B is in agreement with the value measured using Target orientation A.

\medskip
\begin{table}[htbp]
    \centering
    \caption{
        Target Orientation B.
        The statistical and systematic uncertainties on mean energy of neutrons produced by the DD108 neutron generator are shown in columns two and three, respectively.
    }
    \label{tab:target_orientation_b_uncertainties_mean}
    \begin{tabular*}{\columnwidth}{@{\extracolsep{\fill}} lSS}
        \toprule
        Source of Uncertainty & {Statistical} & {Systematic} \\
        & {[\%]} & {[\%]} \\
        \midrule
        n and $\gamma$ peak fits & 0.5 & {-} \\
        Detector position & {-} & 2.4 \\
        Scattering angle & {-} & 0.5 \\
        Choice of bin width & {-} & 0.8 \\
        Choice of fit region & {-} & 0.02 \\
        NaI(Tl) upper area cut & {-} & 1.8 \\
        \midrule
        Total & 0.5 & 3 \\
        \bottomrule
    \end{tabular*}
\end{table}
\medskip

The standard deviation of the underlying neutron energy spectrum was again calculated identically as in Sec.~\ref{sec:tof_orientation_a}.
The standard deviation of the energy distribution of neutrons produced by the DD108 was measured to be 0.067~$\pm$~0.020~(stat)~$\pm$~0.019~(sys)~MeV after subtraction of the gamma ray peak variance.
The uncertainties are listed in Table~\ref{tab:target_orientation_b_uncertainties_sigma}.
This corresponds to a $\sigma/\mu$ of the neutron energies produced by the \dd{} generator of 2.7\%~$\pm$~0.8\%~(stat)~$\pm$~0.8\%~(sys).
The measured width of the neutron energy spectrum produced using Target orientation B is in agreement with the value measured using Target orientation A.

\medskip
\begin{table}[htbp]
    \centering
    \caption{
        Target Orientation B.
        The statistical and systematic uncertainties on the standard deviation of the neutron energy distribution produced by the DD108 neutron generator are shown in columns two and three, respectively.
    }
    \label{tab:target_orientation_b_uncertainties_sigma}
    \begin{tabular*}{\columnwidth}{@{\extracolsep{\fill}} lSS}
        \toprule
        Source of Uncertainty & {Statistical} & {Systematic} \\
        & {[\%]} & {[\%]} \\
        \midrule
        n and $\gamma$ peak fits & 30 & {-} \\
        Detector position & {-} & 2.4 \\
        Scattering angle & {-} & 0.5 \\
        Choice of bin width & {-} & 13 \\
        Choice of fit region & {-} & 0.5 \\
        NaI(Tl) upper area cut & {-} & 25 \\
        \midrule
        Total & 30 & 28 \\
        \bottomrule
    \end{tabular*}
\end{table}
\medskip

\section{Conclusions}

We have proposed a new type of \insitu{} neutron calibration for large dual-phase noble liquid/gas TPCs.
This calibration technique exploits the 3D position reconstruction capabilities of these detectors to measure the scattering angle between multiple interactions in the detector from a single incident neutron and thus absolutely determine the nuclear recoil energy on a per-event basis.
This technique promises to provide a measurement of the charge and light yields of ultra-low-energy nuclear recoils in liquid noble TPCs with reduced experimental uncertainties compared to previous measurements in the field.
This type of \insitu{} neutron calibration can be used to provide a clear confirmation of the WIMP signal response at low masses in the current generation of large TPC dark matter detectors. 

In Sec.~\ref{sec:advanced_dd_techniques}, we described several advanced strategies to enhance the physics reach of the neutron scattering kinematics-based calibration.
First, using the pulsing capabilities of existing commercially available neutron generators to provide a well defined $\mathcal{O}$(10~$\mu$s) neutron bunch width allows for $z$ position reconstruction in the TPC without the traditionally required S1 to provide a $t_{0}$.
This allows the rejection of up to $>$99\% of accidental coincidence-based backgrounds simply from the reduction in duty cycle.
The additional statistical leverage due to the measured number of (no-S1, 1+~S2) events and (1~detected~photon~S1, 1+~S2) events can be used to measure \ly{} lower in energy with a better handle on systematic uncertainties due to threshold effects such as those described in Ref.~\cite{Manalaysay2010}.
Second, a quasi-monoenergetic 272~keV neutron source can be created using a passive deuterium loaded target to reflect neutrons from the \dd{} generator.
This technique provides a $\times$9 reduction in neutron energy.
The reflector neutron source provides an alternative set of calibration systematics and the potential to separate the S1 signals from multiple-scatter vertices due to the $\times$3 reduction in neutron velocity.
These techniques could be exploited in a range of dark matter direct detection experiments including Ge and Si ZIP detectors, semiconductor ionization detectors, noble element single-phase, and dual-phase TPC detectors.

In Sec.~\ref{sec:brown_dd_neutron_energy_spectrum_measurement}, we measured the neutron energy spectrum of an Adelphi Technology, Inc.\ DD108 neutron generator at 90$^{\circ}$ relative to the deuterium ion beam direction.
We characterized the outgoing neutron energy spectrum in two directions relative to the asymmetrical neutron production surface to determine the optimal orientation for the proposed nuclear recoil calibration. 
In both cases, the measured mean neutron energy is in agreement with the theoretical expectation of 2.45~MeV for this experimental setup.
We also report the intrinsic width ($\sigma/\mu$) of the outgoing neutron energy distribution.
The width of the distribution of neutron energies for Target Orientation A and Target Orientation B was measured to be 4.4\%~$\pm$~0.6\%~(stat)~$\pm$~0.8\%~(sys) and 2.8\%~$\pm$~0.8\%~(stat)~$\pm$~0.8\%~(sys), respectively.
The measured mean and width of the neutron energy distribution for Target Orientation A and Target Orientation B are in agreement within quoted uncertainties, indicating negligible dependence on rotation along the azimuthal direction.
This characterization of the DD108 neutron spectrum confirms the suitability of the device for the calibrations described in Sec.~\ref{sec:dd_proposal} and Sec.~\ref{sec:advanced_dd_techniques}.

\section*{Acknowledgments}

This work was partially supported by the U.S. Department of Energy (DOE) under award number DE-SC0010010.
This research was conducted using computational resources and services at the Center for Computation and Visualization, Brown University.
We thank Adelphi Technology, Inc.\ for providing data on the Brown DD108 hardware, and for lending their expertise during neutron generator operation.

\bibliography{dd108_spectrum_paper}

\end{document}